\documentclass[acmsmall]{acmart}
\renewcommand\footnotetextcopyrightpermission[1]{} 
\usepackage{titlesec}  

\titleformat{\subsubsection}
  {\normalfont\bfseries}
  {\thesubsubsection}
  {1em}
  {}
\titlespacing*{\subsubsection}
  {0pt}
  {3.25ex plus 1ex minus .2ex}
  {2ex plus .2ex}

\AtBeginDocument{%
  }
    
\usepackage{algorithm}
\usepackage{algorithmic}

\usepackage{utfsym}

\usepackage{tabularx} 
\usepackage{array}

\usepackage{multirow} 
\usepackage{booktabs} 
\usepackage{makecell} 
\usepackage{array} 

\usepackage{ragged2e} 

\usepackage{caption}
\usepackage{array} 
\usepackage{cellspace} 
\usepackage{array} 
\usepackage{cellspace} 
\usepackage{colortbl} 
\definecolor{lightgray}{gray}{0.8} 

\usepackage[utf8]{inputenc}
\usepackage{listings}
\usepackage{xcolor}

\lstdefinelanguage{Solidity}{
    keywords={contract, function, public, payable, mapping, address, uint256, msg, sender}, 
    keywordstyle=\color{cyan!80!black}\bfseries, 
    morestring=[b]", 
    stringstyle=\color{orange!80!black}, 
    morecomment=[l]{//}, 
    morecomment=[s]{/*}{*/}, 
    commentstyle=\color{gray!60}\itshape, 
    sensitive=true 
}

\begin{document}

\title{AI-Based Vulnerability Analysis of NFT Smart Contracts}

\author{Xin Wang}
\email{xinwang@hainanu.edu.cn}
\affiliation{%
  \institution{Hainan University}
  \city{Haikou}
  \country{China}
}
\author{Xiaoqi Li}
\email{csxqli@ieee.org}
\affiliation{%
 \institution{Hainan University}
  \city{Haikou}
  \country{China}
}

\begin{abstract}
With the rapid growth of the NFT market, the security of smart contracts has become crucial. However, existing AI-based detection models for NFT contract vulnerabilities remain limited due to their complexity, while traditional manual methods are time-consuming and costly. This study proposes an AI-driven approach to detect vulnerabilities in NFT smart contracts.

We collected 16,527 public smart contract codes, classifying them into five vulnerability categories: Risky Mutable Proxy, ERC-721 Reentrancy, Unlimited Minting, Missing Requirements, and Public Burn. Python-processed data was structured into training/test sets. Using the CART algorithm with Gini coefficient evaluation, we built initial decision trees for feature extraction. A random forest model was implemented to improve robustness through random data/feature sampling and multitree integration. GridSearch hyperparameter tuning further optimized the model, with 3D visualizations demonstrating parameter impacts on vulnerability detection.

Results show the random forest model excels in detecting all five vulnerabilities. For example, it identifies Risky Mutable Proxy by analyzing authorization mechanisms and state modifications, while ERC-721 Reentrancy detection relies on external call locations and lock mechanisms. The ensemble approach effectively reduces single-tree overfitting, with stable performance improvements after parameter tuning. This method provides an efficient technical solution for automated NFT contract detection and lays groundwork for scaling AI applications.

\end{abstract}


\maketitle
\section{INTRODUCTION}
NFT (Non-fungible Token), or non-fungible token, is a certified storage unit operating on a unique and indivisible blockchain platform. Currently, it is mostly in the form of electronic files of the source files of artistic creation, and its value is reflected by virtual cryptocurrencies\citep{kong2024characterizing, huang2023security}.

The current focus on the NFT smart contract market revolves mainly around four aspects: classification and mining of vulnerabilities, development of audit tools, repair of vulnerability strategy, and construction of the NFT ecosystem\citep{liu2025sok}. Due to the complexity of NFT smart contract vulnerabilities, there is currently no mature large-scale artificial intelligence detection model specifically for NFT smart contract vulnerabilities in the market.\citep{li2024cobra} Therefore, the analysis, detection, repair, and maintenance of vulnerabilities in the NFT smart contract require a significant amount of resources, which is not conducive to the continuous healthy development of the existing NFT market. \citep{gandhi2025ai}

Therefore, this paper proposes an AI-based NFT smart contract vulnerability analysis project in response to the lack of large-scale artificial intelligence analysis models. The goal is to analyze vulnerabilities and carefully analyze the generated models, starting from data, models, and effects, and conducting a large amount of demonstration and experimental work \citep{li2024detecting}. We strive to accumulate experience in NFT smart contract vulnerabilities and contribute to the application of large-scale artificial intelligence models. \citep{praitheeshan2019security}

There are already several examples of defect detection in AI-based smart contracts,rocessing (NLP) and machine learning algorithms to perform static analysis and vulnerability detection on smart contract code\citep{bu2025enhancing}, identifying potential security risks and providing improvement suggestions, offering reliable security assurance for blockchain developers;\citep{niu2024unveiling} Li Tao and others proposed a public audit of smart contracts based on game theory; Chuang Ma1 proposed HGAT,\citep{ma2023hgat} a hierarchical graph attention network-based detection model.
Internationally, the Harvard University Blockchain Security Laboratory has developed a static analysis tool for smart contracts using artificial intelligence technology, which can automatically detect vulnerabilities and security risks in smart contracts; \citep{sendner2023smarter} and Anzhelika Mezina \citep{mezina2023detecting} and others proposed a method combining binary classification and multiclassification to detect vulnerability in smart contracts in their paper. 

This paper focuses on the currently widely used NFT smart contracts, conducting a comprehensive and in-depth study on their related security issues, collecting a large number of applied NFT smart contract codes; after careful sorting and differentiation, the attacks are divided into five categories; they are processed and studied separately and finally trained, \citep{elghalhoud2022data} in addition, this paper also continues to explore its shortcomings and possible future work directions and priorities \citep{vsutas2024automated}, providing its own opinions for the improvement of NFT smart contract detection technology.

The paper has the following contribution:
\begin{itemize}
\item \textbf{  Sufficient and accurate data: }This paper compiled a dataset of 16,527 smart contracts, and analyzed and labeled the vulnerabilities in their line codes. We also carefully cleaned and preprocessed the data to exclude inaccurate or low-quality code samples, ensuring that the data quality on which the model is based is highly reliable. \citep{ali2023metaverse}
\item \textbf{Targeted solutions: We conducted an in-depth analysis of NFT smart contracts and discovered five main defects:} Risky Mutable Proxy, ERC-721 Reentrancy, Unlimited Minting, Missing Requirements, and Public Burn. These defects may seriously impact the contract's security and robustness. To solve these problems, we formulate targeted solutions to improve the program's quality and robustness. \citep{yang2023definition}
\item \textbf{Sufficient training and organization:} In our research, we performed a detailed parameter tuning for the random forest model. Adjusting the number of trees, the depth of trees, the selection of features, and other parameters, we explored many different training results. This sufficient training and organization process improved our model and provided more reliable prediction and evaluation capabilities for smart contract code security. \citep{chen2024improving}
\end{itemize}

\section{BACKGROUND}
\subsection{Smart Contracts}
Smart contracts, in terms of classification, belong to protocols. Their primary application scenario is within the blockchain, containing a large number of code functions. \citep{li2025scalm}. Additionally, they can interact and operate with other smart contracts to achieve a series of required functionalities. \citep{li2023overview}Similar to protocols, they need to follow specified steps and processes for application. Moreover, smart contracts allow two parties to conduct trusted transactions independently without the need for a traditional trusted center. These transactions are traceable and irreversible \citep{gong2023scgformer}. When a specific scenario or action triggers the corresponding terms of a smart contract in a certain way, the smart contract code can execute accordingly.
\begin{figure*}[t]
    \centering
    \includegraphics[width=0.9\linewidth]{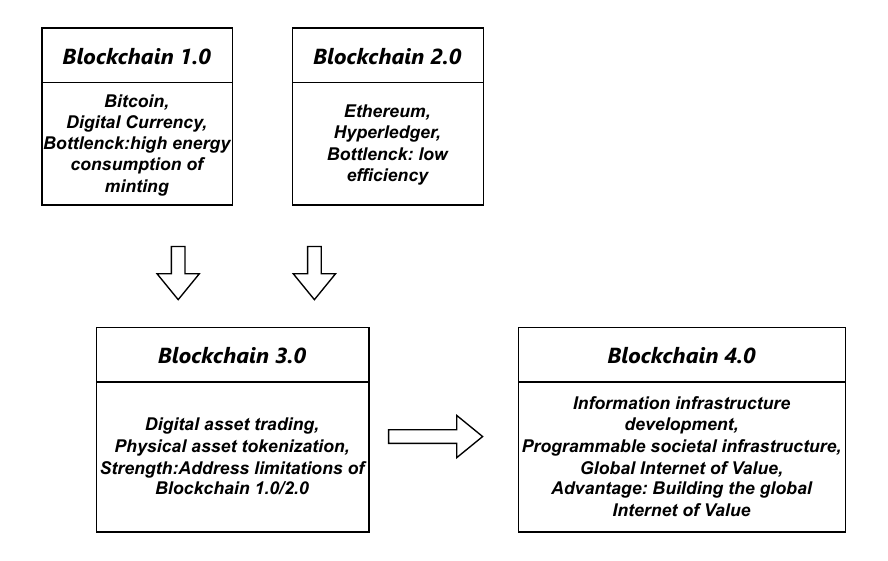} 
        \caption[Figure 1]{Blockchain Industry Development Trends}
    \label{framework}
\end{figure*}

As one of the important components of blockchain technology, smart contracts were first proposed by the renowned cryptographer Nick Szabo in 1994 \citep{lakshminarayana2022towards}. However, due to technological and infrastructure constraints, they have not been fully implemented. Although smart contracts are now widely used on the Internet in areas such as automatic payments and drone sales, they are mostly limited to contracts between individuals and institutions. \citep{li2020characterizing}The main reason is the increasing unfamiliarity between people in modern society, making it difficult to establish precise and effective constraint mechanisms, thus resulting in higher living costs for most people when dealing with issues. Using blockchain technology, trust issues between people can be resolved through technical methods, promoting the further development of smart contracts.
\begin{figure*}[t]
    \centering
    \includegraphics[width=0.92\linewidth]{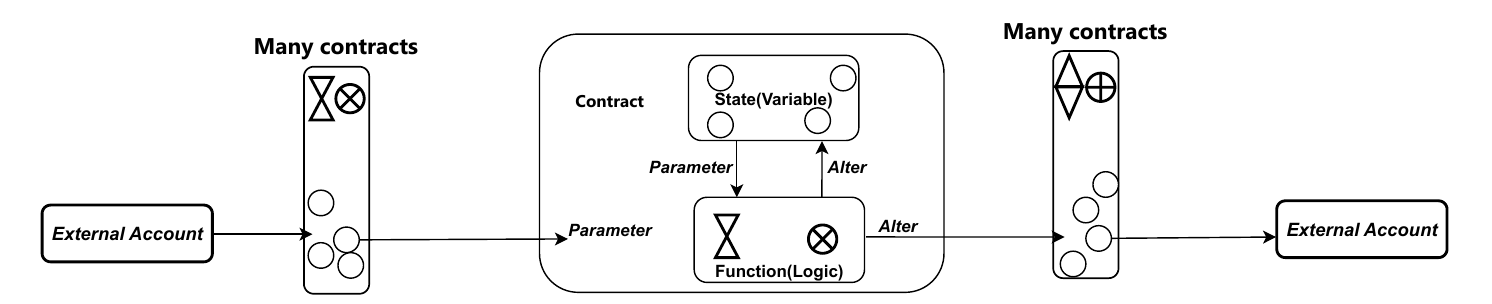} 
        \caption[Figure 1]{Smart Contract Logic}
    \label{framework}
\end{figure*}

Blockchain, with its decentralized nature and the characteristic that data cannot be altered at certain points, theoretically and technically solves the trust issues between two or more unfamiliar parties, laying the foundation for the large-scale application of smart contracts. \citep{li2024stateguard}The birth of Ethereum marked the beginning of this phase, applying smart contract technology on the blockchain and endowing Ethereum with more functionalities and application possibilities.
Smart contracts are gradually becoming one of the core technologies of blockchain, with their importance increasingly prominent. With the advancement of China's BRICS and Belt and Road initiatives, blockchain technology, characterized by decentralization and data immutability, has more realistic and extensive application scenarios!

From a narrow perspective, blockchain technology is a distributed ledger based on chronological iteration. Each block is connected end to end, forming a chain-like structure. During operation, its security is ensured through cryptographic principles, such as timestamps.\citep{liu2024gastrace}. From a broad perspective, blockchain uses a transmission and verification structure as its architecture, solving data processing issues through consensus mechanisms. It achieves a decentralized infrastructure and distributed computing paradigm by designing programmable smart contracts \citep{kim2021automotive}. It is both an architecture and a paradigm.

In blockchain, except for the first block, each block contains the transaction data and verification data (such as timestamps) of the previous block. In transactions, a Merkle tree is used to obtain hash values, ensuring security. However, it should be noted that if certain individuals or organizations control more than 50\% of the computational power of the blockchain, they can manipulate the content of the blockchain. Furthermore, if the initial written content is incorrect, blockchain technology makes it difficult to forge and alter the incorrect content \citep{jung2019fair}.

In current blockchain explanations, we divide the blockchain structure into six layers: data layer, consensus layer, network layer, incentive layer, application layer, and contract layer. The first three are core layers, and the latter three are extension layers.

In real life, based on differentiated scenarios and user needs, \citep{li2024scla}we set different nodes and access mechanisms, providing multiple choices divided into public chains, private chains, and consortium chains.

\subsection{Random Forest Model}
Random forest is a type of ensemble learning, that expands on decision trees and integrates the advantages of a large number of decision trees \citep{han2021estimating}.
Decision trees mainly handle classification and regression problems, classifying based on one feature and then proceeding until no further division is possible \citep{van1993decision}.

Random forest is a type of ensemble learning that approximates the best result by constructing a specified number of multiple decision trees. Since each decision tree is independent and trained on different sample sets obtained by resampling the training data, each decision tree is trained on a random subset of the original data set \citep{tian2020crown}. Below is an example of an ensemble learning mechanism.

In addition to random sampling of training data, random forests introduce other randomness. During the construction of each decision tree, only a random subset of characteristics is considered to divide, reducing the excessive influence of individual characteristics on predictions and increasing the diversity of the model.\citep{li2024defitail} This introduction of randomness helps to reduce overfitting and gives random forests a better generalization ability. Randomness is crucial to the success of the forest \citep{freeman2016random}. Below is a specific demonstration of randomness.

\begin{figure*}[t]
    \centering
    \includegraphics[width=0.92\linewidth]{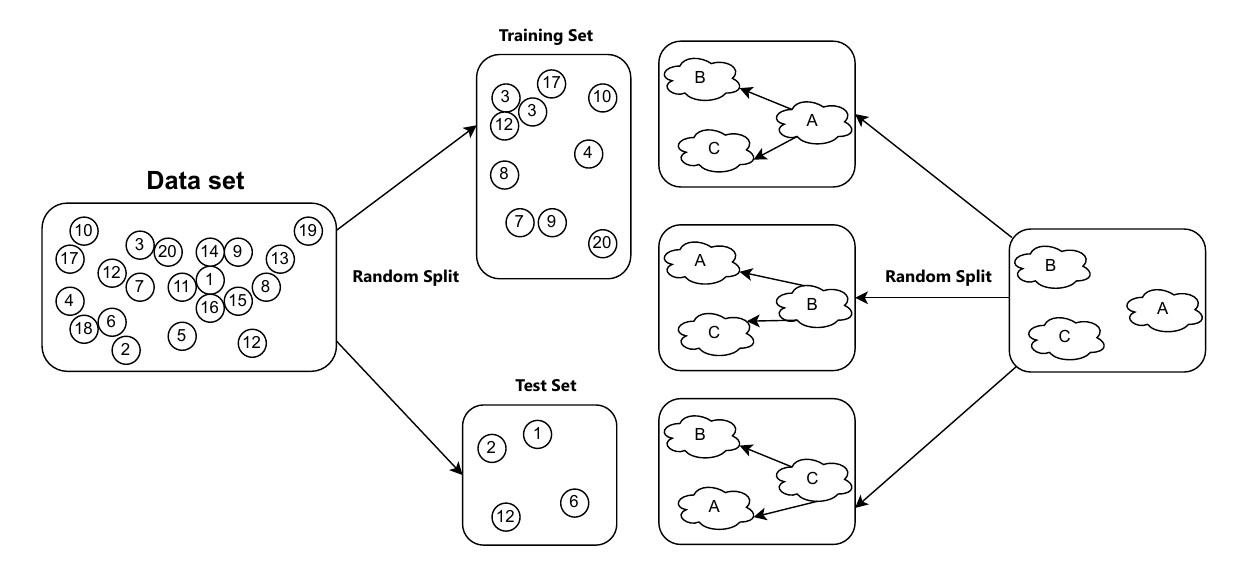} 
        \caption[Figure 3]{Randomness Demonstration Diagram}
    \label{framework}
\end{figure*}
In random forests, when we need to predict new datasets, to ensure accuracy, each decision tree independently predicts the sample without interference from other decision trees. For classification problems, the final prediction result can be determined through a one-vote-per-tree voting process, with the majority vote determining the best result. For regression problems, we sum all predicted values and average them to eliminate errors, ultimately obtaining the result. \citep{boruah2023transparent}

In summary, the random forest model is based on the construction of multiple decision trees using random data sampling and random feature selection methods. Through continuous iteration, it integrates the prediction results of each decision tree, thereby improving the accuracy and adaptability of the model. \citep{chen2024improving}

\section{PRELIMINARY PREPARATION}
\subsection{Data Processing}
Based on our research and screening, we ultimately selected the smart code files provided by Shuo Yang in his paper "Definition and Detection of Defects in NFT Smart Contracts" as our original dataset. This dataset contains a total of 16,527 smart code files.

Facing a large number of smart contract code files for the convenience of subsequent labeling (the labeled content is placed in CSV files). Since the sorting method in Windows is different from that in Excel, we wrote functions to modify the file names accordingly.
Remove non-English parts from the SOL files to ensure correct labeling.

Finally, we perform data classification and labeling. During the data processing, through research and analysis, we identified and categorized five corresponding issues: Risky Mutable Proxy, ERC-721 Reentrancy, Unlimited Minting, Missing Requirements, and Public Burn.
We label the sorted CSV files, marking 1 for files with the issue and 0 for those without.

\subsection{Vulnerability Analysis}
\begin{itemize}
\item \textbf{Risky Mutable Proxy:}When a proxy contract is used in a smart contract to manage administrator permissions, attackers may exploit code vulnerabilities or improper parameter settings in the contract to gain control of the proxy contract or tamper with the contract's state\citep{li2024guardians}, leading to instability and security issues in the contract.

\item  \textbf{ERC-721 Reentrancy:}
The ERC-721 Reentrancy vulnerability is a common security issue in NFT smart contracts compatible with the ERC-721 standard. This vulnerability is similar to the general reentrancy attack principle, which may result in the theft of funds or tampering with the contract state. This vulnerability is usually associated with the transfer function in smart contracts, where attackers exploit inconsistencies in the contract state to repeatedly call other functions during the transfer execution \citep{wang2024smart}, leading to repeated transfers of funds or state tampering.

\item  \textbf{Unlimited Minting:}
The Unlimited Minting vulnerability is a potential security risk in NFT smart contracts, allowing malicious users to mint new tokens without limit, causing the token supply to exceed the expected or designed range. This vulnerability may arise from the improper implementation of the minting function in smart contracts.

\item  \textbf{Missing Requirements:}
The Missing Requirements vulnerability is a potential security risk in NFT smart contracts, where the underlying logic fails to meet or implement specific functional or security requirements [15]. When running smart contracts, the absence of necessary protective measures may lead to various issues.

\item  \textbf{Public Burn:}
The Public Burn vulnerability is a common issue in the processing of smart contracts [16]. It refers to the situation where, during the processing of smart contracts, sometimes it is necessary to destroy some currency or processes, but the corresponding mechanisms and defensive measures are not properly established. During processing, many illegal operations may go unnoticed by the smart contract, leading to unnecessary damage and trouble. For example, repeatedly destroying a currency can cause logical confusion.
\end{itemize}

\subsection{Vulnerability Examples}
\begin{itemize}

\item \textbf{Risky Mutable Proxy Explanation:}
The owner variable is used to store the address of the contract owner.
The current proxy variable is used to store the proxy address that is currently authorized.
The setProxy function sets a new proxy address as the currently authorized proxy. Only the contract owner can call this function.
The transferFrom function is used to transfer NFTs from one smart contract address to another. Only the currently authorized proxy address can call this function.

\item \textbf{Risky Mutable Proxy analysis:}
In the contract, only the contract owner can call the setProxy function to change the current proxy address. If an attacker can control the contract owner's address, or if the contract owner carelessly grants control to another address, the attacker can call the setProxy function to set a malicious proxy address as the current proxy.
Once a malicious proxy address is set as the current proxy, the attacker can call the transferFrom function to transfer NFTs to any address without the NFT owner's control. In this case, the proxy address change can occur at runtime, hence the term mutable proxy vulnerability. The attacker exploits the contract's permission change functionality to bypass the original permission control, leading to unauthorized NFT transfers.

\begin{figure*}[t]
    \centering
    \includegraphics[width=0.8\linewidth]{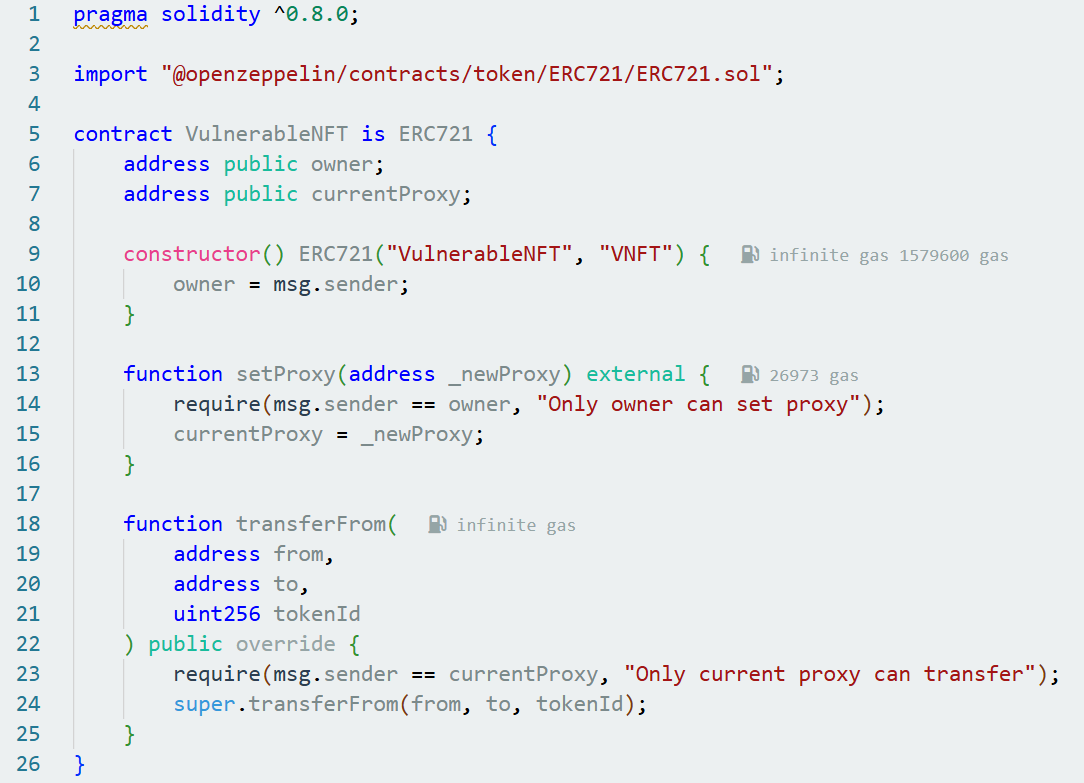} 
        \caption[Figure 4]{Risky Mutable Proxy Vulnerability}
    \label{framework}
\end{figure*}

\end{itemize}
\begin{itemize}
\item \textbf{ERC-721 Reentrancy Explanation:}
Here, the mint function primarily handles scheduling issues in the process, allocating NFTs after generating them.
The transfer function generally transfers NFT addresses from one address to another.
The withdrawal function is used mainly for payment issues during user operations. It checks if the balance is greater than 0 and if the payment can be completed. If so, it proceeds with the payment transaction.
\item \textbf{ERC-721 Reentrancy Analysis:}
Attackers can exploit this vulnerability to create a malicious contract that repeatedly calls the transfer and withdrawal functions to extract the contract's balance. This is because, in the current contract, the balance update operation occurs before the transfer, creating a vulnerability that attackers can exploit. To prevent Reentrancy vulnerabilities, measures such as executing the transfer operation before the balance update or using Solidity's reentrancyGuard modifier can be taken.

\begin{figure*}[t]
    \centering
    \includegraphics[width=0.76\linewidth]{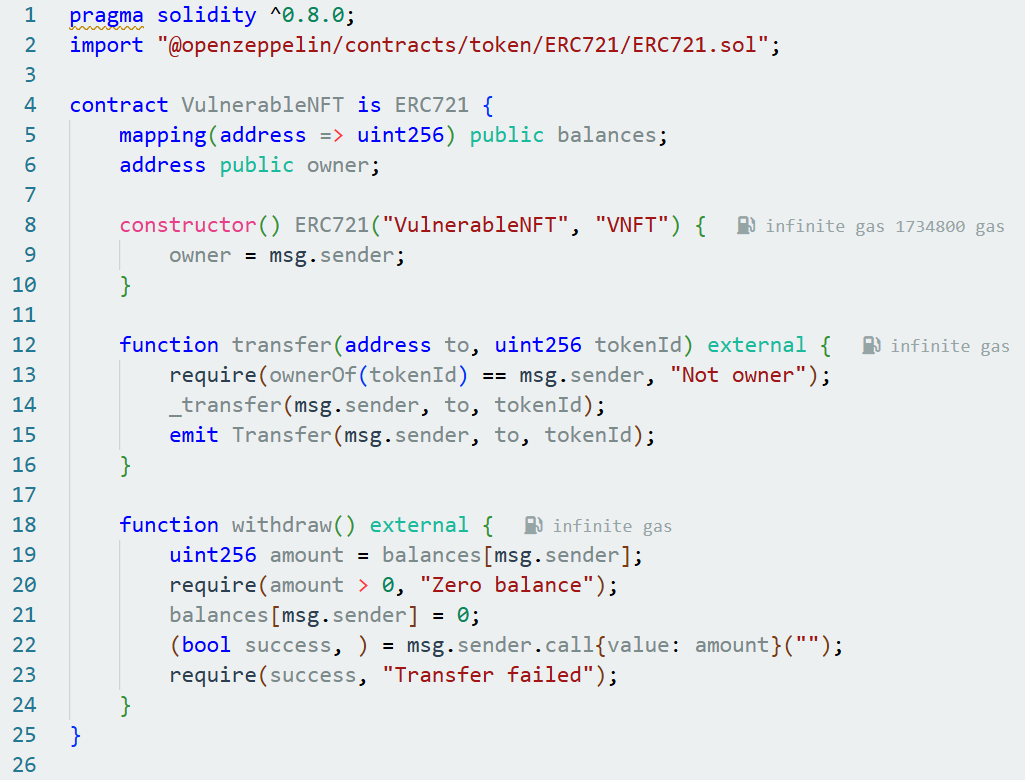} 
        \caption[Figure 4]{ERC-721 Reentrancy Vulnerability}
    \label{framework}
\end{figure*}
\end{itemize}
\begin{itemize}
\item \textbf{Unlimited Minting Explanation:}
totSupply is mainly used to check how many NFT tokens the user currently owns.
Since the mint function allows anyone to mint tokens without restrictions, if not modified, it can lead to uncontrollable increases in token supply, affecting basic blockchain operations.
\item \textbf{Vulnerability analysis:}
In this example, we define a contract named UnlimitedMiningNFT, which inherits from OpenZeppelin's ERC721 contract. The contract has a mint function that allows anyone to mint new NFTs and assign them to the caller. However, this contract has a serious vulnerability: there are no restrictions on the number of tokens that can be minted.
\begin{figure*}[t]
    \centering
    \includegraphics[width=0.8\linewidth]{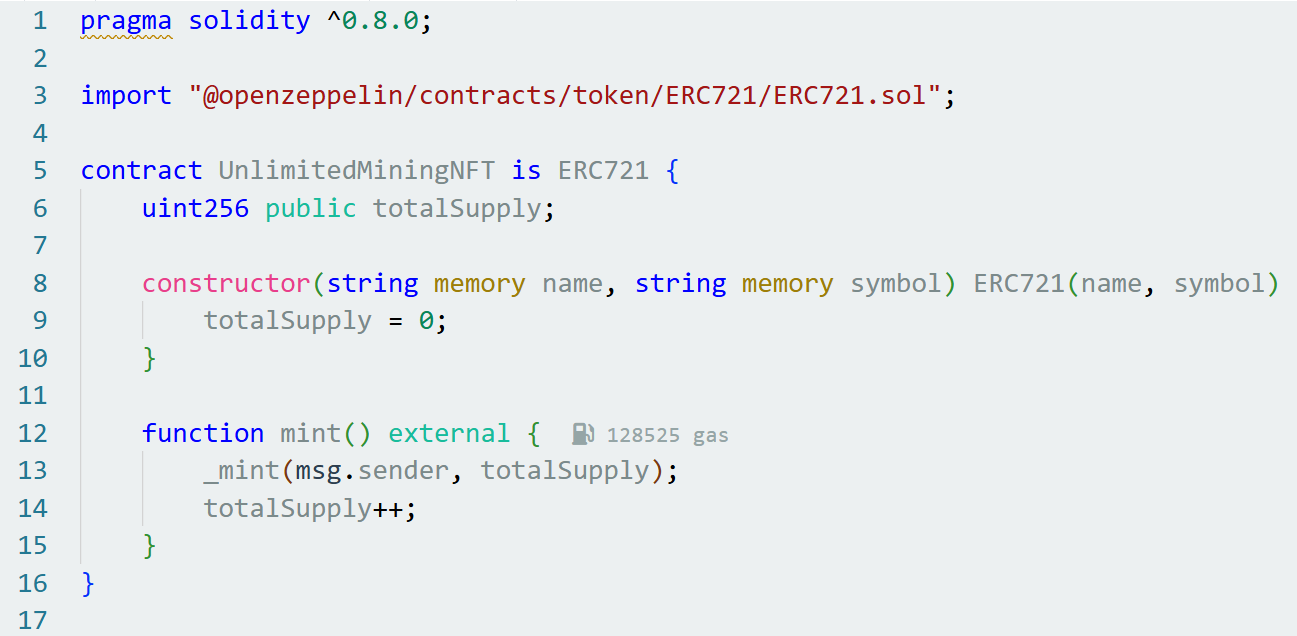} 
        \caption[Figure 5]{Unlimited Minting Vulnerability}
    \label{framework}
\end{figure*}
\end{itemize}
\begin{itemize}
\item \textbf{Missing Requirements Explanation:}
During the processing of the smart contract, we did not control the mint function, allowing many people to create NFT tokens.

\item \textbf{Missing Requirements Impact:}
Without a professional control mechanism to limit the creation of NFT tokens, a large number of tokens are created. When tokens increase uncontrollably, the corresponding tokens in the market become worthless, leading to inflation and market issues.
\begin{figure*}[t]
    \centering
    \includegraphics[width=0.8\linewidth]{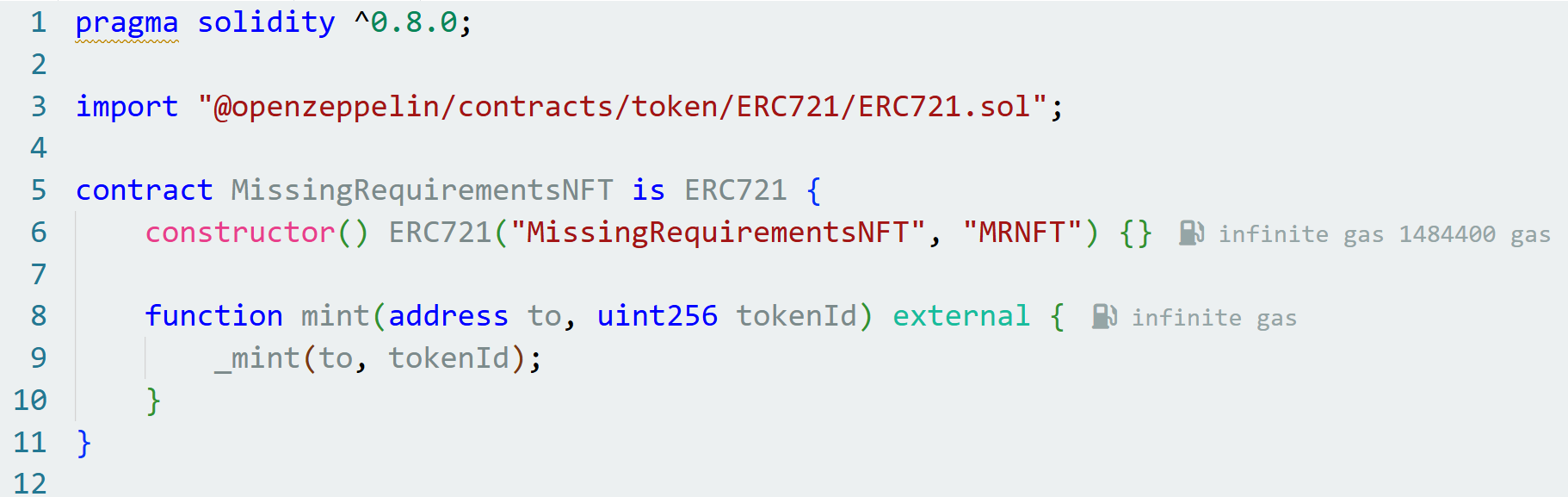} 
        \caption[Figure 6]{Missing Requirements Vulnerability}
    \label{framework}
\end{figure*}
\end{itemize}
\begin{itemize}

\item \textbf{Public Burn Explanation:}
The burn function in the contract is publicly callable, allowing anyone to call it to destroy specified NFTs. Since there are no restrictions to check if the caller has the right to destroy the specified NFT, anyone can destroy any NFT at will.

\item \textbf{Public Burn Impact:}
Without proper permission controls, anyone can destroy any NFT at will, leading to the irreversible loss of NFT ownership. Attackers can exploit this vulnerability to damage the market value of NFTs or affect the interests of NFT holders.
If the NFTs in the contract have actual value or represent real assets, the public burn function may lead to financial losses. A lack of necessary permission controls can make the contract vulnerable to malicious attacks or abuse.
\begin{figure*}[t]
    \centering
    \includegraphics[width=0.8\linewidth]{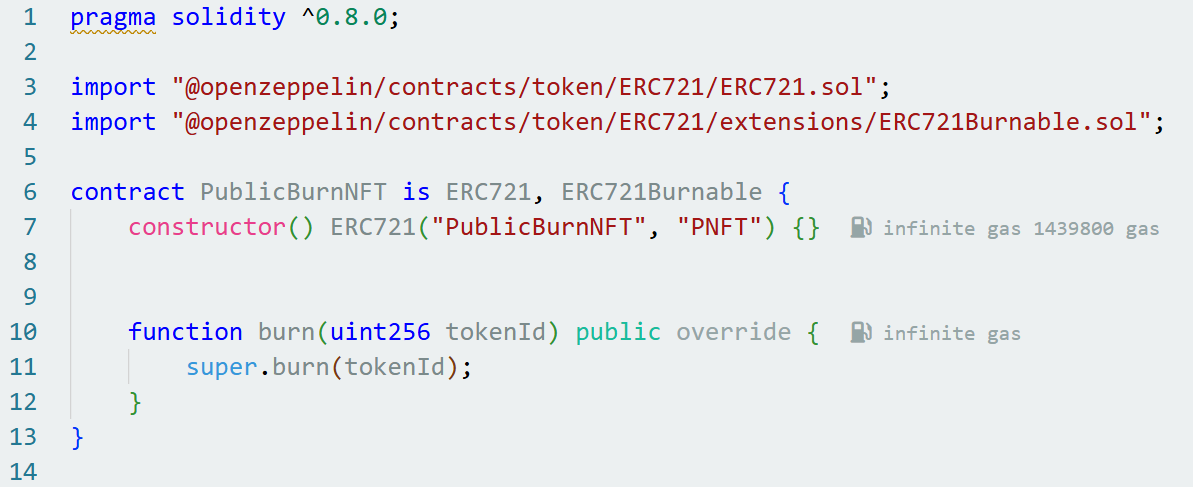} 
        \caption[Figure 6]{Public Burn Vulnerability}
    \label{framework}
\end{figure*}
\end{itemize}

\section{CONSTRUCTION OF DECISION TREES}
This section mainly focuses on the construction of decision trees, serving as the foundation for building the random forest model. The process includes feature extraction, algorithm selection, decision tree generation, and pruning for continuous improvement.

We will address the five vulnerabilities listed above, perform feature extraction, handle training, and finally, summarize the analysis and application for general models.\citep{loh2011classification}
\subsection{Feature Extraction}
\begin{itemize}

\item \textbf{Risk Mutable Proxy Feature Extraction:}For the risk of Mutable proxy vulnerability, feature extraction is conducted from six aspects: whether there is a proxy contract call, whether there is a parameter setting function, whether there is permission control, whether there is a state change record, whether there is a risk operation function, and whether there is condition detection. \citep{utgoff1989incremental}
\item \textbf{ERC-721 Reentrancy Feature Extraction:}For the ERC-721 reentrancy vulnerability, feature extraction is performed from eight aspects: whether there is an external call, whether there is a locking mechanism in the contract, whether the contract state is not properly handled, whether there is an asset transfer vulnerability in the contract, whether the contract state modification is separated, whether the contract does not properly handle the return value of external contract calls, whether the contract lacks necessary state checks and updates, and whether the contract does not properly handle exceptions. \citep{salzberg1994c4}
\item \textbf{Unlimited Mining Feature Extraction:}For unlimited mining vulnerability, feature extraction is conducted from eight aspects: whether the minting function lacks permission verification, whether there is a lack of total supply limit, whether there is a lack of condition checks, whether there is an extensible minting logic, whether there is a lack of a triggerable stop mechanism in the smart contract, whether there is an unreasonable minting fee in the smart contract, whether there are external dependency calls in the smart contract code, and whether there is a possibility of unlimited minting due to permission or role abuse in the contract. \citep{luu2016making}
\item \textbf{Missing Requirement Feature Extraction:}For the Missing Requirement vulnerability, feature extraction is conducted from eight aspects: whether there is a function definition in the contract but lacks an input validation function, whether there is a lack of security checks, whether there is a lack of transfer restriction functions, whether there is a lack of auditing and readability functions, whether there is a lack of event triggering, whether there is a lack of permission control, whether there is an upgrade mechanism, and whether there is a lack of asset metadata validation. \citep{tsankov2018securify}
\item \textbf{Public Burn Feature Extraction:}For the public Burn vulnerability, feature extraction is conducted from six aspects: whether there is a lack of authentication, whether there is a lack of confirmation or recovery mechanism, whether there is a mismatched authorization mechanism, whether the owner identity is not considered, whether there is a lack of event logs, and whether there is a duplicate destruction mechanism. \citep{surucu2022survey}
\end{itemize}
\subsection{Feature Selection and Division}

There are three decision tree algorithms: ID3, C4.5, and CART. After consideration and practice, we chose the CART algorithm for the following reasons:

\begin{itemize}
\item The ID3 algorithm lacks a pruning step for the generated decision tree, which can easily lead to overfitting \citep{lalduhsaka2022anomaly,ghaleb2020effective}.

\item The C4.5 algorithm requires sorting numerical category attributes during tree construction, which requires ensuring that the generated data can be stored in the host memory's dataset. When the provided data volume is too large, the generated data volume becomes too large, making it difficult to run the program.

\item The C4.5 algorithm generates a multi-branch tree, which requires more complex processing and more resources and time during operation. The CART algorithm, being a binary tree, consumes fewer resources and requires less time \citep{hasmin2019penerapan}.

\item The CART algorithm uses the Gini criterion for judgment during training data processing, as it does not require logarithmic operations that consume a lot of resources \citep{chen2023risk}. Given the large data volume in this study, it is more suitable. \citep{chen2020survey}
\end{itemize}

Classification and Regression Tree has two main functions: handling classification problems and regression problems. Depending on the situation, the processing method differs.  

Generally, when the dependent variable of the data is discrete, we use classification methods for processing. During each judgment and classification, the category with the highest probability is selected as the predicted category for that point. However, when the dependent variable of the data is continuous, classification cannot be used for division (if divided, all points would form all categories, losing their original meaning). We use regression methods for processing, taking the average of all predicted results to obtain the predicted value.

When handling problems, the CART algorithm generates a binary tree, meaning each classification results in only two situations. If more than two results appear for the same feature, it would cause a logical error (in this paper, a feature is divided into False and True, so this situation does not need to be considered).

\subsection{Gini Coefficient}
The entropy model consumes a lot of resources during operation because it handles a large amount of logarithmic-level operations. The Gini index, on the other hand, simplifies the complexity of the model while retaining the high accuracy of the entropy model. The Gini index represents the impurity of the model; the smaller the Gini coefficient, the lower the impurity\citep{kumar2013construction}, and the better the effect (when judging whether smart contract code has a vulnerability, it fits the concept of purity), the better the feature.

\[Gini(D) = \sum_{k=1}^{K} \left[ \frac{|C_k|}{|D|} \left( 1 - \frac{|C_k|}{|D|} \right) \right] 
         = 1 - \sum_{k=1}^{K} \left( \frac{|C_k|}{|D|} \right)^{\!2}
\]
\[
Gini(D|A) = \sum_{i=1}^{n} \frac{|D_i|}{|D|} \cdot Gini(D_i)
\]
\begin{itemize}

 \item k represents the category
 \item D represents the sample set

 \item $C^k$ represents the subset of samples in set D that belong to the kth category
\end{itemize}
The meaning of the Gini index is: randomly selecting two samples from the training data, the probability that they are classified into different classes by the model. The smaller it is, the higher the purity, and the better the effect. The Gini index can be used to measure whether the data distribution is balanced, with values ranging from 0 to 1 represents 100\% effect, fully meeting the requirements. 1 represents 0\% effect, completely unequal.

In this study, to divide whether it belongs to the vulnerability, it is a binary classification in CART, and the formula can be simplified to

\[
\text{Gini}(D_1) = 1 - \left( \frac{|C_1|}{|D|} \right)^2
\]
\[
\text{Gini}(D_2) = 1 - \left( \frac{|C_1|}{|D|} \right)^2 - \left( \frac{|C_2|}{|D|} \right)^2
\]
\[
\text{Gini}(D|A) = \frac{|D_1|}{|D|} \cdot \text{Gini}(D_1) + \frac{|D_2|}{|D|} \cdot \text{Gini}(D_2)
\]
Where represent D, $D_1$, $D_2$ the number of samples in datasets D, $D_1$, and $D_2$, respectively.

\subsection{Generating CART Decision Trees}
\begin{itemize}
   
\item \textbf{Risk Variable Proxy Decision Tree Generation:}
This vulnerability involves six functions. For ease of writing during operation and simplicity in decision tree generation, we assign them serial numbers A1-A6. They are authentication, recovery mechanism, owner identity check, event logs, and duplicate destruction issues.
As shown in the table below:
\begin{table}
    \centering
    \begin{tabular}{lll}
        \hline
        \textbf{Function Name}   & \textbf{Feature Number} \\
        \hline
        detect\_proxy\_call    &A1 \\
        detect\_parameter\_setting     &A2  \\
        detect\_permission\_control       &A3 \\
        detect\_state\_change	    &A4 \\
        detect\_insurance\_function            &A5 \\
        detect\_condition\_check            &A6\\
        \hline
    \end{tabular}
    \caption{Risk Mutable Proxy Function Feature Comparison
}
    \label{tab:plain}
\end{table}

The partial sample data table obtained after feature calculation is shown below:
\begin{table}[h]
    \centering
    \setlength{\abovecaptionskip}{0.cm}
    
    \resizebox{0.8\linewidth}{!}{
    \begin{tabular}{cccccccc}
    \toprule
        \textbf{File} &\textbf{A1} & \textbf{A2}  & \textbf{A3} &\textbf{A4} &\textbf{A5} &\textbf{A6} &\textbf{Risk}  \\  \hline
        addcfaaabdbcbfccf.sol &Flase  &Flase  & True & True & Flase & Flase &1 \\
        bdbdbbcabdc.sol &Flase  &True  &Flase & Flase & Flase & Flase &0\\
        Bccffcaccbcf.sol &Flase  &Flase  & Flase & Flase & Flase & Flase &0 \\
        Acdbaafcbabcbs.sol &Flase  &True  & Flase & Flase & Flase & Flase &0 \\
        Feaddbbbcdfacd.sol &Flase  &Flase  & Flase & Flase & True & True &1 \\
        Dfefadedbae.sol &Flase  &Flase  & Flase &Flase & Flase & True &0 \\
        \hline
    \end{tabular}}
    \caption{Partial Sample Data Table}
    \label{performance_measures}
\end{table}

Where the value of the feature return is False and True, False represents the absence of the feature, True represents the presence of the feature, 0 represents no risk, and 1 represents risk. Calculate the Gini coefficient for each feature value and select the optimal feature and the optimal split point. After sorting, the following table is obtained.

\begin{table}
    \centering
    \begin{tabular}{lll}
        \hline
        \textbf{Function}   & \textbf{Gini Index} \\
        \hline
        A1    &0.17 \\
        A2     &0.42  \\
        A3       &0.15 \\
        A4	    &0.39 \\
        A5            &0.34 \\
        A6            &0.28\\
        \hline
    \end{tabular}
    \caption{Feature Gini Coefficient Comparison
}
    \label{tab:plain}
\end{table}
From the above calculation, Gini(D1, A1)=0.17 is the smallest, so it is selected as the root node, and the recursion continues. The decision tree is established as follows:
\begin{figure*}[t]
    \centering
    \includegraphics[width=0.6\linewidth]{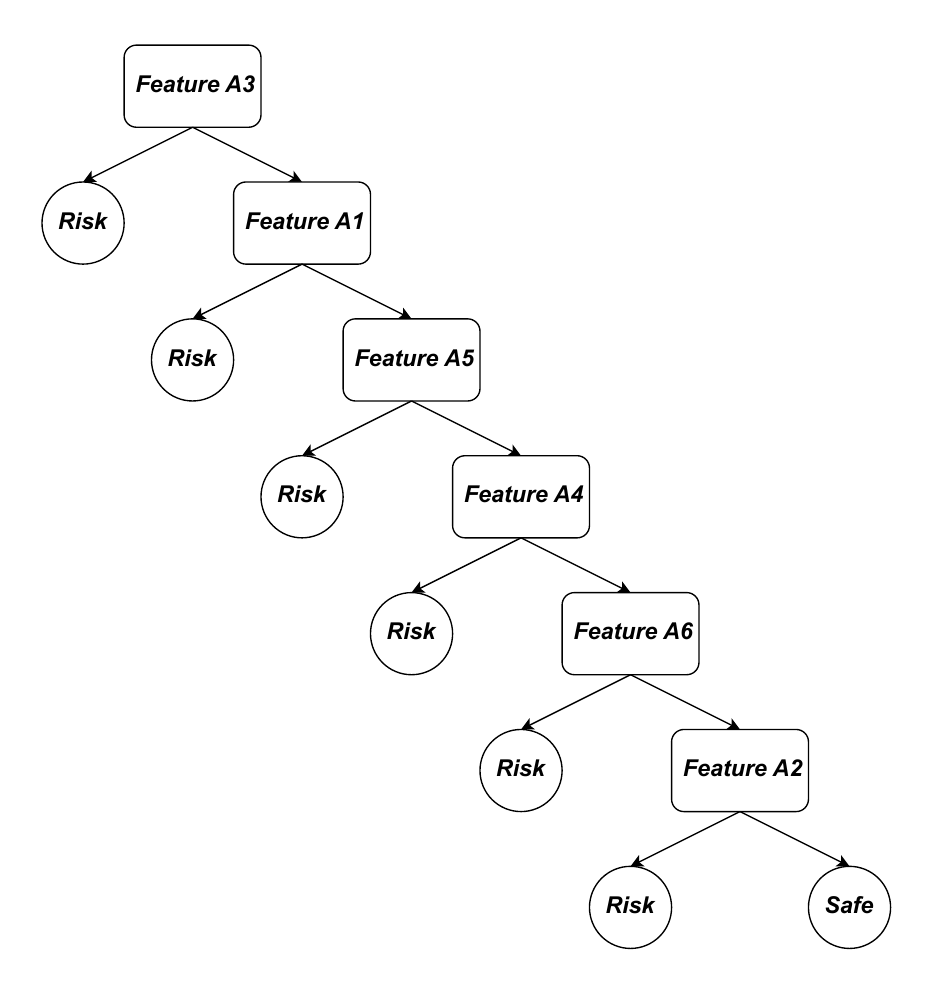} 
        \caption[Figure 9]{Risk Variable Proxy Decision Tree}
    \label{framework}
\end{figure*}
From top to bottom, whether there is an authorization mechanism, authentication, event logs, owner identity check, duplicate destruction, or recovery mechanism is determined.

\item \textbf{ERC-721 Reentrancy Decision Tree Generation:}
The above shows the process of generating the risk variable proxy decision tree. The generation process of other decision trees is similar, so it is not repeated here. Only the feature corresponding numbers and the generated decision trees are listed.
\begin{table}
    \centering
    \begin{tabular}{lll}
        \hline
        \textbf{Function Name}   & \textbf{Feature Number} \\
        \hline
        detect\_external\_call\_locations    &B1 \\
        detect\_locking\_mechanism     &B2  \\
        detect\_reentrancy\_vulnerability       &B3 \\
        detect\_asset\_transfer\_vulnerability	    &B4 \\
        detect\_state\_change\_separation           &B5 \\
        detect\_unhandled\_external\_call         &B6\\
        detect\_missing\_state\_check\_update            &B7\\
        detect\_missing\_exception\_handling           &B8\\
        \hline
    \end{tabular}
    \caption{ERC-721 Reentrancy Function Feature Comparison
}
    \label{tab:plain}
\end{table}

\begin{figure*}[t]
    \centering
    \includegraphics[width=0.6\linewidth]{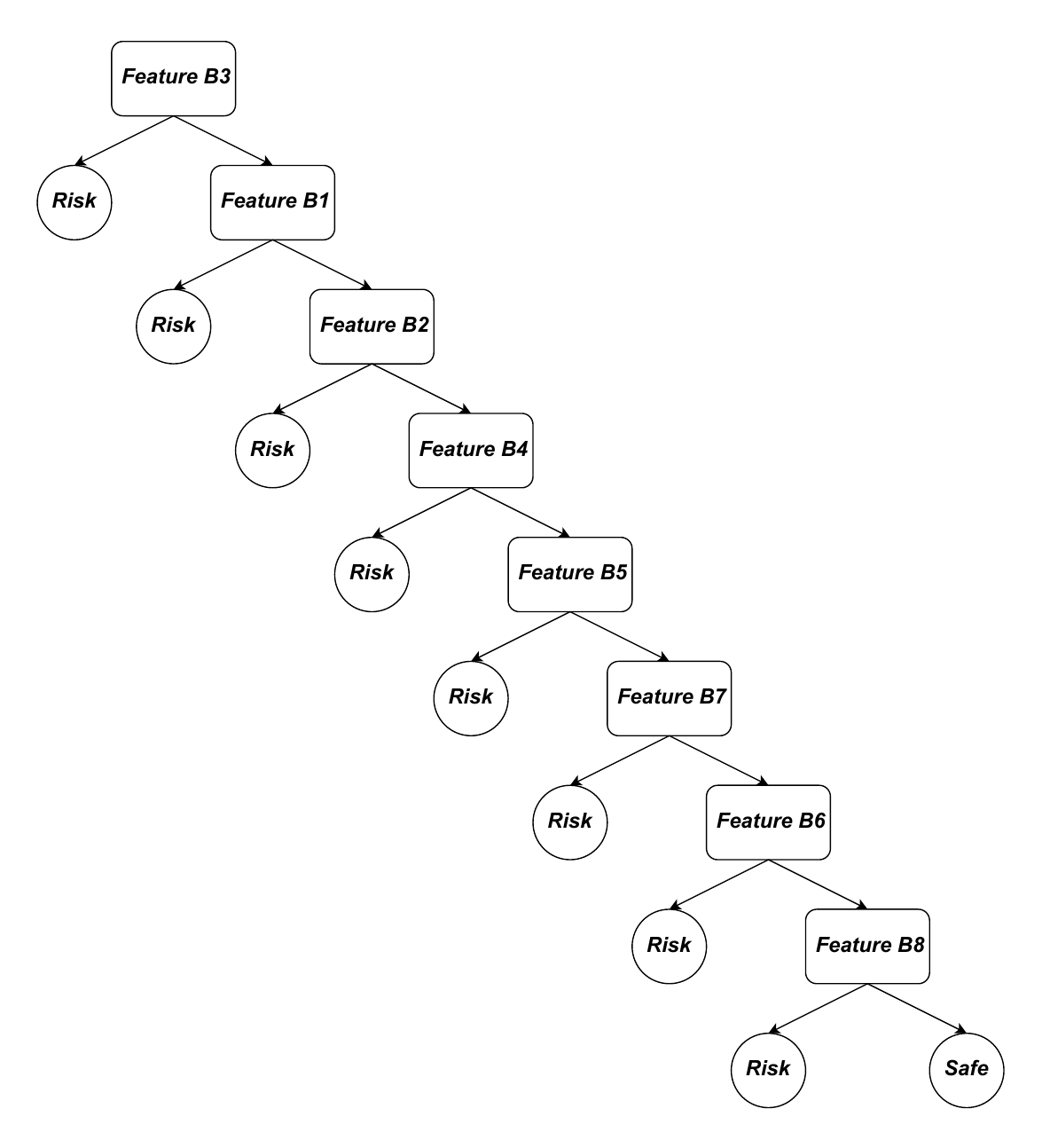} 
        \caption[Figure 10]{ERC-721 Reentrancy Decision Tree}
    \label{framework}
\end{figure*}
From top to bottom, it is whether the state is not properly handled, whether there is an external call, whether there is a locking mechanism, whether there is an asset transfer vulnerability, whether there is state modification separation, whether there is a lack of necessary state checks, whether the return value of external contract calls is properly handled, and whether general exception vulnerabilities are properly handled.

\item \textbf{Unlimited Mining Decision Tree Generation:}From top to bottom, it is whether there is an extensible minting logic, whether there is a lack of total supply limit, whether there is a lack of a triggerable stop mechanism, whether there is a lack of condition checks, whether there is a lack of permission restrictions, whether there is an unreasonable minting logic, whether there are external dependency calls, and whether there is permission abuse.
\begin{table}
    \centering
    \begin{tabular}{lll}
        \hline
        \textbf{Function Name}   & \textbf{Feature Number} \\
        \hline
        detect\_unverified\_minting    &C1 \\
        detect\_total\_supply\_limit     &C2  \\
        detect\_condition\_missing     &C3 \\
        detect\_extendable\_minting\_logic	    &C4 \\
        detect\_is\_unlimited\_minting           &C5 \\
        detect\_unreasonable\_minting\_fee         &C6\\
        detect\_external\_calls            &C7\\
        detect\_permission\_role\_abuse          &C8\\
        \hline
    \end{tabular}
    \caption{Unlimited Mining Function Feature Comparison
}
    \label{tab:plain}
\end{table}

\begin{figure*}[t]
    \centering
    \includegraphics[width=0.6\linewidth]{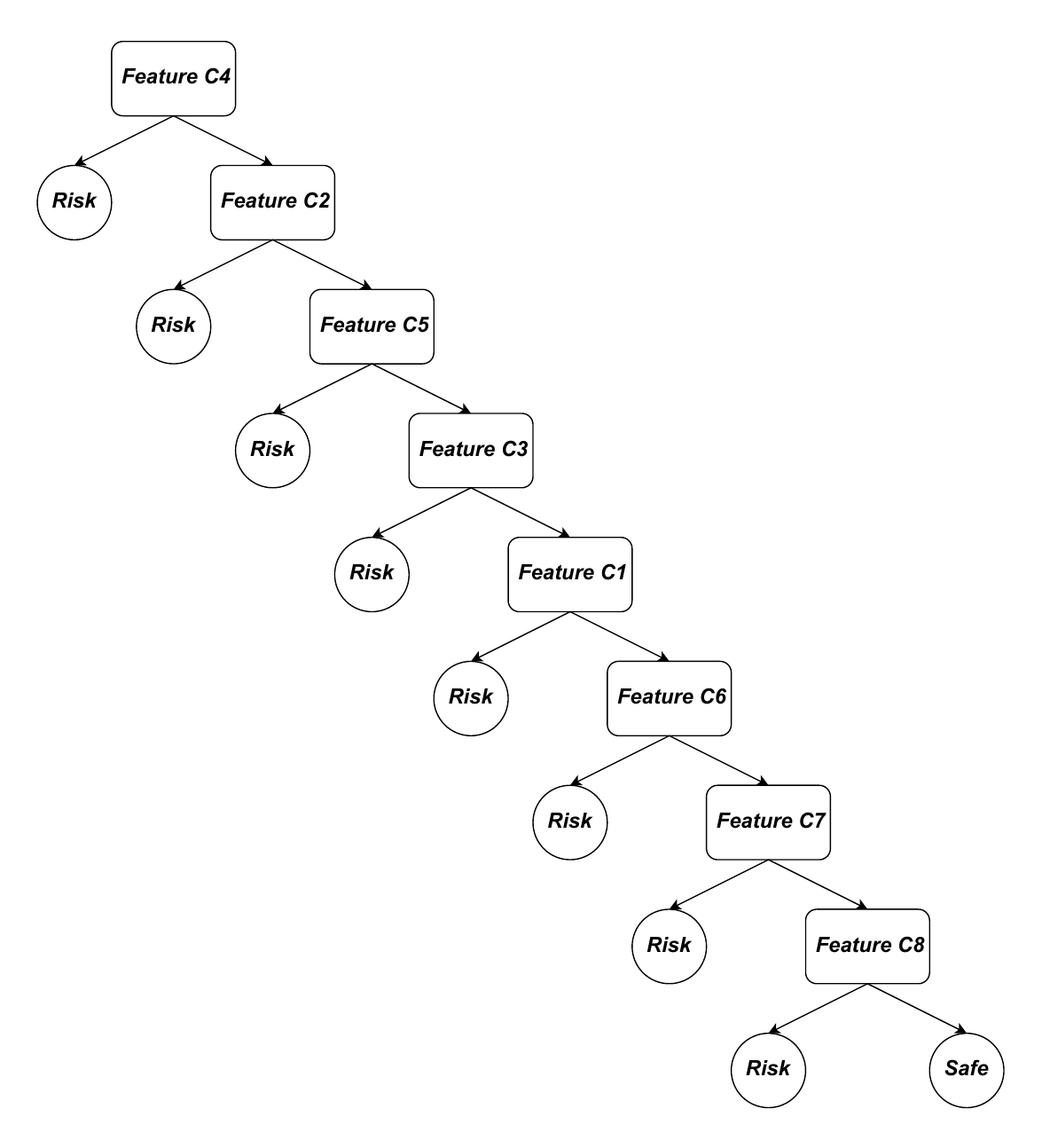} 
        \caption[Figure 11]{Unlimited Mining Decision Tree}
    \label{framework}
\end{figure*}

\item \textbf{Ignored Requirement Decision Tree Generation:}From top to bottom, it is whether there is an upgrade mechanism, whether there is a lack of transfer restriction functions, whether there is a lack of event triggering, whether there is a lack of asset metadata validation, whether there is a function definition but lacks an input validation function, whether there is a lack of security checks, whether there is a lack of auditing and readability functions, and whether there is a lack of event triggering.
\begin{table}
    \centering
    \begin{tabular}{lll}
        \hline
        \textbf{Function Name}   & \textbf{Feature Number} \\
        \hline
        detect\_missing\_input\_validation    &D1 \\
        detect\_missing\_security\_checks    &D2  \\
        detect\_missing\_transfer\_restrictions     &D3 \\
        detect\_missing\_auditing\_functions	    &D4 \\
        detect\_missing\_event\_functions           &D5 \\
        detect\_missing\_permission\_functions         &D6\\
        detect\_missing\_update\_mechanism            &D7\\
        detect\_missing\_metadata\_validation          &D8\\
        \hline
    \end{tabular}
    \caption{Missing Requirement Function Feature Comparison
}
    \label{tab:plain}
\end{table}

\begin{figure*}[t]
    \centering
    \includegraphics[width=0.6\linewidth]{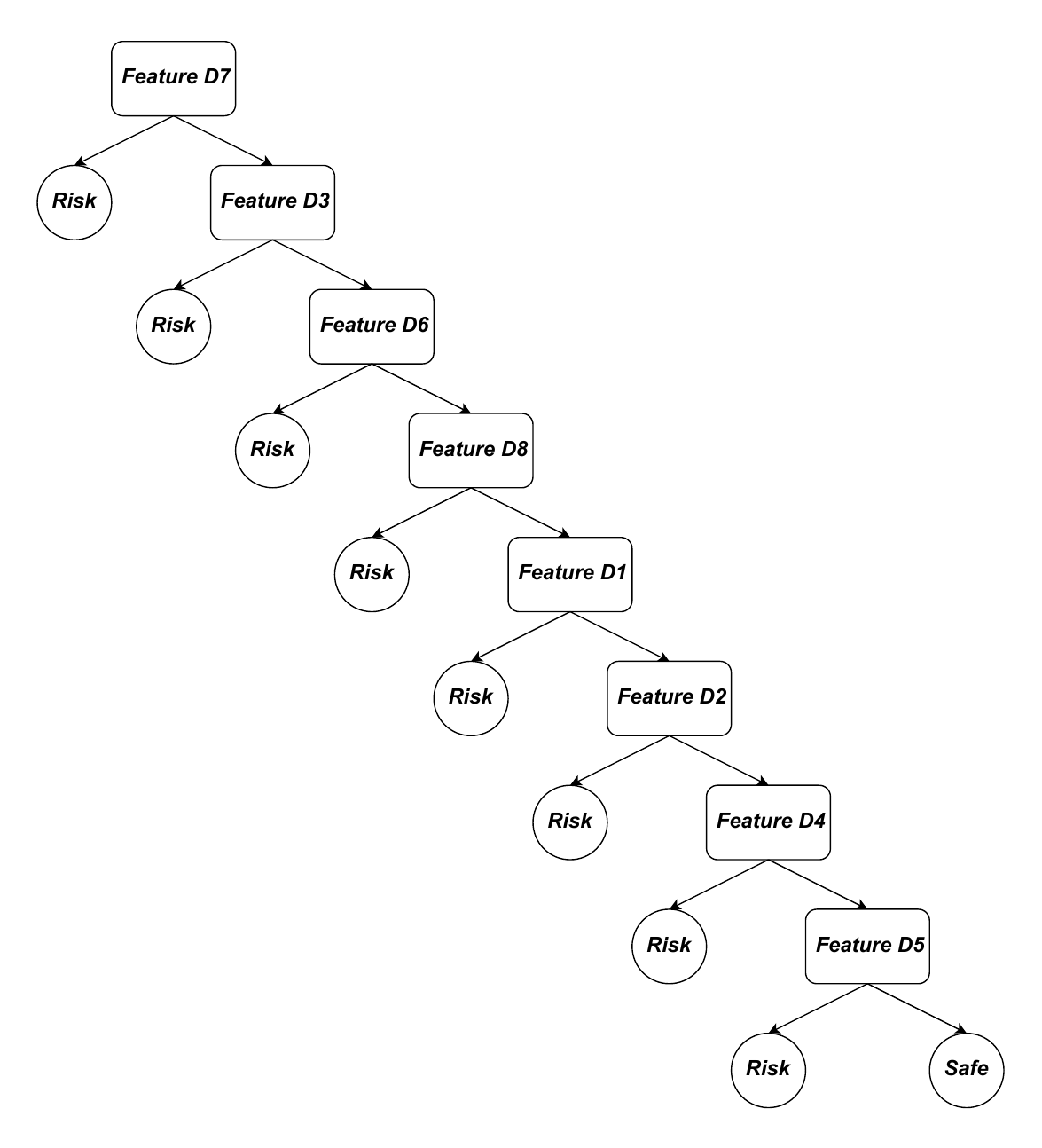} 
        \caption[Figure 12]{Missing Requirement Function Feature Comparison}
    \label{framework}
\end{figure*}

\item \textbf{Public Burn Decision Tree Generation:}
\begin{table}
    \centering
    \begin{tabular}{lll}
        \hline
        \textbf{Function Name}   & \textbf{Feature Number} \\
        \hline
        detect\_burn\_requires\_authentication    &E1 \\
        detect\_lack\_of\_confirmation\_recovery    &E2  \\
        detect\_improper\_authorization     &E3 \\
        detect\_unverified\_owner	    &E4 \\
        detect\_missing\_event\_logs           &E5 \\
        detect\_duplicate\_destruction         &E6\\
        \hline
    \end{tabular}
    \caption{Public Burn Function Feature Comparison
}
    \label{tab:plain}
\end{table}

\begin{figure*}[t]
    \centering
    \includegraphics[width=0.6\linewidth]{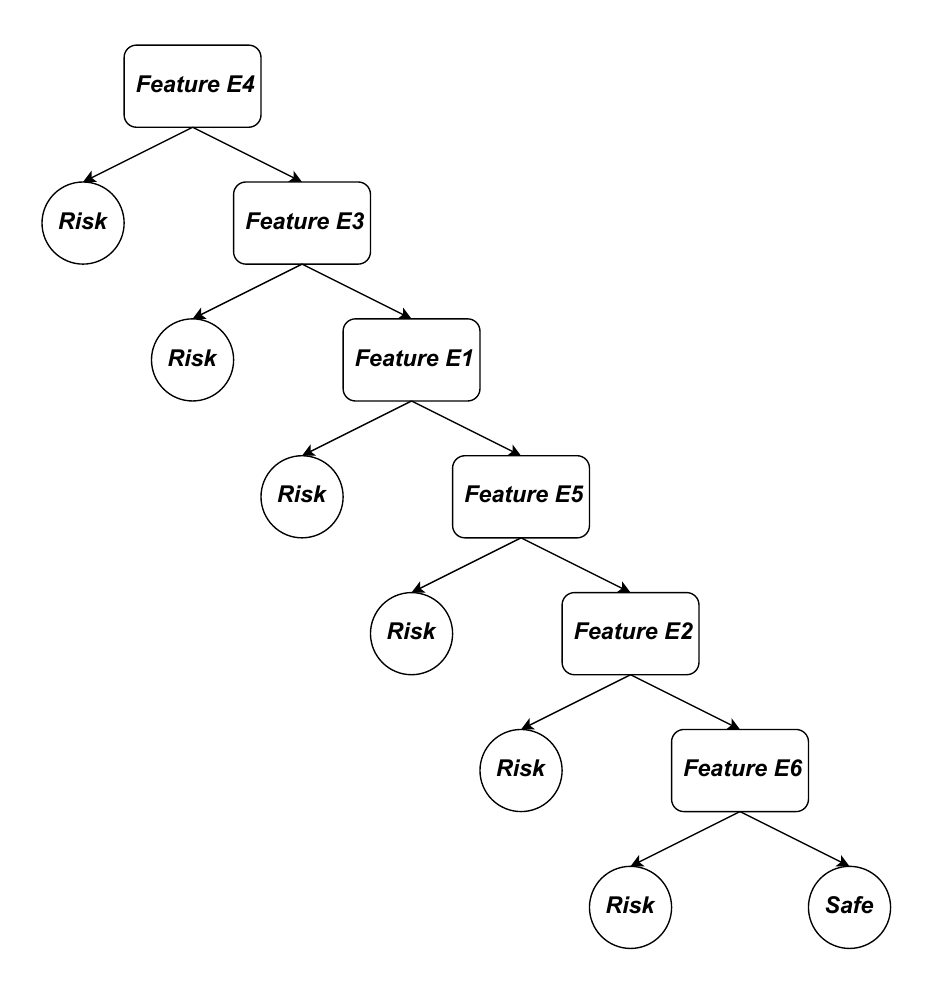} 
        \caption[Figure 13]{Public Burn Decision Tree}
    \label{framework}
\end{figure*}

From top to bottom, it is whether the owner identity is not considered, whether there is a mismatched authorization mechanism, whether there is a lack of authentication, whether there is a lack of event logs, whether there is a lack of confirmation or recovery mechanism, and whether there is a duplicate destruction mechanism.
\end{itemize}
\subsection{Decision Tree Pruning}
Since decision tree algorithms may overfit the training set \citep{boruah2023transparent}, leading to poor generalization ability, to improve the usability of the code, the generated decision tree needs to be appropriately reduced. The CART algorithm first organizes and generates the decision tree, then prunes and performs cross-validation, selecting the most accurate and adaptable solution.

The algorithm mainly consists of two aspects: First, starting from the bottom node of the decision tree, continuously reduce and iterate until the last node, forming a non-repeating subsequence.

Then, using cross-validation methods, test the generated decision tree sequence on the validation dataset, selecting the best-performing CART decision tree.

During pruning, the loss function is as follows:
\[ C_{\alpha} = C(T) + \alpha |T| \]
\begin{table*}
    \centering
    \begin{tabular}{lll}
        \hline
        \textbf{Symbol}   & \textbf{Meaning} \\
        \hline
        T    &Any subtree \\
        C(T)     &Prediction error of the data  \\
        |T|       &Number of leaf nodes in the subtree \\
        \multirow{2}{*}{$\alpha$} 	    &Regularization parameter, balancing the  \\ &fitting degree of  training data and model complexity \\
        \hline
    \end{tabular}
    \caption{Feature Gini Coefficient Comparison
}
    \label{tab:plain}
\end{table*}
\begin{itemize}
\item When $\alpha$=0, there is no regularization, meaning the original generated CART decision tree is the optimal solution.

\item When $\alpha$=$+\infty$, the regularization degree is very high, meaning the decision tree containing only the root node of the CART tree is the best-performing subtree.
Generally, the larger $\alpha$ is, the more thorough the pruning, and the better the effect.

\item Using a recursive method, starting from zero, $\alpha$ increases sequentially, 0<$\alpha_{0}$<$\alpha_{1}$<$\ldots$<$\alpha_{n-1}$<$\alpha_{n}$<$+\infty$,

forming [$\alpha_{i}$,$\alpha_{(i+1)}$), i=0,1,2$\ldots$n, a total of n+1 intervals. The subsequence obtained through pruning corresponds to each interval from small to large \citep{kumar2013fuzzy}.
\end{itemize}
\hspace{1em}Starting from a decision tree $T_{0}$, for any internal feature node t of $T_{0}$, the loss function is                     
\[ C_{\alpha}(t) = C(t) + \alpha\]

The loss function of the subtree $T_{t}$ with t as the root node is
\[ C_{\alpha} = C(t) + \alpha\ |T| \]

When $\alpha=0$ or $\alpha$ $\to$ $+0$
\[C_{\alpha}(T_{t})<C_{\alpha}(t)\]

When $\alpha$ increases to a certain extent, there will be
\[C_{\alpha}(T_{t})=C_{\alpha}(t)\]

When $\alpha$ continues to increase
\[C_{\alpha}(T_{t})>C_{\alpha}(t)\]

At this point, $T_{t}$ and t have the same loss function, but since t has fewer nodes than $T_{t}$.

We solve
\[C_{\alpha}(T_{t})=C_{\alpha}(t)\]
\hspace{1em}to get 
\[g(t)=\frac{C(T)-C(T_{t})}{|T_{t}-1|}\]
\hspace{1em}Thus, we can calculate the value $\alpha$ for each internal node t in the complete decision tree $T_{0}$.
\[g(t)=\frac{C(T)-C(T_{t})}{|T_{t}-1|}\]

In this paper, g(t) represents the degree of reduction in the overall loss function after pruning. For example: in $T_{0}$, pruning the $T_{t}$ with the smallest g(t) value, the resulting subtree is $T_{1}$, and this g(t) value is set as $\alpha_{1}$. We get $T_{1}$ as the optimal subtree for the interval [$\alpha_{1}$,$\alpha_{2}$).

Then, iterate until the root node, forming a sequence of \{ $T_{0}$,$T_{1}$,$\ldots$,$T_{n}$ \}. Using the Gini index criterion mentioned in 3.3, test the subtree sequence on the new validation set, select the best-performing subtree, and output it as the optimal decision tree.

\section{RANDOM FORESTS MODELS}
\subsection{Introduction to Ensemble Learning}
The Random Forest model, as an ensemble learning algorithm, is based on weak classifiers. When dealing with classification and regression problems, \citep{surucu2022survey} the final results are processed through voting and averaging methods \citep{pawar2023text}, ensuring the accuracy and adaptability of the overall model. Due to its excellent stability, it is widely used in various business scenarios. \citep{sagi2018ensemble}

The outstanding performance of RF is largely attributed to its key features: randomness and the forest. Randomness effectively solves the overfitting problem, while the forest structure avoids many adverse situations, ensuring greater accuracy. The model is primarily composed of the following concepts. Since it is built on decision trees through ensemble learning, we will provide a supplementary introduction to ensemble learning below.

Ensemble learning is not a specific step or algorithm but rather a concept. We can use the stories "Many hands make light work" and "Three Cobblers with their wits combined surpass Zhuge Liang" as examples. Ensemble learning leverages the principle of "many hands make light work." It does not create something new like the cobblers but integrates existing algorithms to improve accuracy \citep{yu2010developing, hossain2023novel}.
In terms of completing tasks, the approach can be divided into three categories: Stacking, Boosting, and Bootstrap Aggregating (Bagging).
\begin{itemize}

\item \textbf{Stacking:}
Stacking uses the results generated by individual learners as input to train a secondary learner, iteratively generating the model.
The basic idea of stacking is to integrate the results of multiple learners to form new feature variables. These new features, along with the original features, are input into the secondary learner for training \citep{li2024structural}. This allows the secondary learner to utilize the predictions of the base learners, resulting in better predictive performance. 
The general process of stacking is as follows:

\item Split the data: Divide the original data into training and testing parts.

\item Train base learners: Train multiple different base learners, such as decision trees, support vector machines, and neural networks.

\item Generate new features: Use the base learners to predict the training and test sets, using the prediction results for each sample as new features.

\item Train the secondary learner: Combine the original features with the new features and train the secondary learner.

\item Predict: Use the completed model to predict the untested portion of the data.
\end{itemize}
The advantage of stacking is that it fully utilizes the strengths of each base learner. Compared to simple averaging or weighted averaging, stacking typically yields better performance. However, stacking also has some drawbacks, such as requiring more computational resources and a more complex tuning process.

\begin{itemize}
\item\textbf{Bagging:}
Bagging (Bootstrap Aggregating) uses the bootstrap method to draw a large number of samples from the original dataset for training. After training, the samples are returned, and multiple independent base learners are trained using these subsamples. Finally, their prediction results are processed to generate the final ensemble model. The core idea of bagging is voting \citep{rout2018ensemble}. Each model has equal influence, and the final answer is determined by voting \citep{boruah2023transparent}. Typically, the results obtained through bagging have a smaller variance.

Bagging constructs multiple slightly different subsamples through the above process. These subsamples are then used to train multiple base learners. Predictions are made using these slightly different subsamples. Since the samples and learners are different, these predictions are independent and unaffected by other factors, providing good adaptability for the model.The general process of bagging is as follows:

\begin{figure*}[t]
    \centering
    \includegraphics[width=0.6\linewidth]{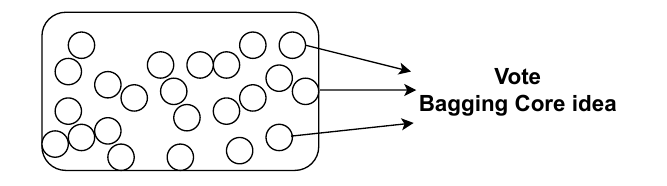} 
        \caption[Figure 14]{Bagging Core Idea}
    \label{framework}
\end{figure*}

 \item Use the bootstrap method to draw multiple subsamples from the original data and return them after training.

 \item Train the subsamples to obtain multiple independent learners.

 \item During prediction, use different learners to predict the test samples separately, and average or vote based on the prediction results.

 \item Finally, integrate the results produced by the base learners to obtain the final result.
Bagging has advantages in reducing variance. For general learning algorithms, we can perform parallel computations during the process to train multiple base learners. It is more effective for learners with high variance.
\end{itemize}
\hspace{1em}Through bagging, the model's adaptability to problems is improved. Many studies and practices can be enhanced through this process, ensuring its effectiveness. Figure 15 is a demonstration of the bagging approach.
\begin{figure*}[t]
    \centering
    \includegraphics[width=0.6\linewidth]{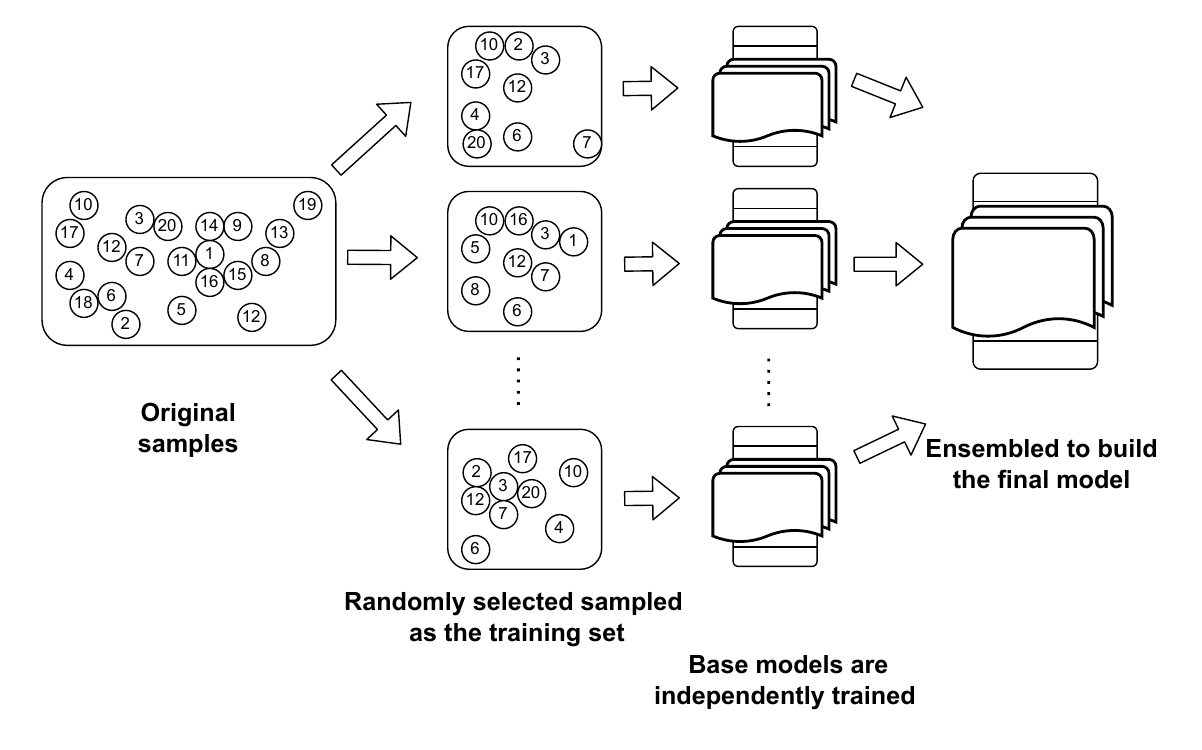} 
        \caption[Figure 15]{Bagging Specific Idea Diagram}
    \label{framework}
\end{figure*}

\begin{itemize}
\item\textbf{Boosting:}
Boosting is a category of ensemble learning methods, known as Boosting in English. It is based on multiple weak learners, integrated in a certain way to produce an efficient learner.

The main idea of boosting is to appropriately weight the performance of the data in each round of operation. In each iteration, the learner's weights are readjusted. The data is processed again with incomplete resampling, allowing the data to be retrained, and then their weights are updated based on their performance \citep{nakach2022hybrid}. Through this method, the model reduces the influence of problematic learners, thereby improving overall performance. \citep{chiu2022applying} The general process of boosting is as follows:

\begin{figure*}[t]
    \centering
    \includegraphics[width=0.6\linewidth]{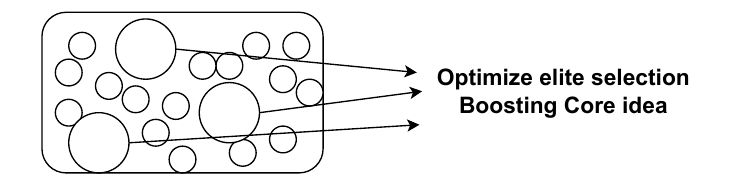} 
        \caption[Figure 16]{Boosting Core Idea}
    \label{framework}
\end{figure*}

\item \textbf{Uniform weights:} Upon receiving the data required by the model, we ensure that each sample has the same weight to maintain consistent initial data influence.

\item \textbf{Repeated training:} A large amount of data is fed into the learner, processed, and weights are redistributed based on their influence.

\item \textbf{Combination:} The results obtained from repeated iterative training are combined, typically prioritizing learners with higher weights.

\item \textbf{Prediction:} The model is validated using the validation set.
\end{itemize}
\hspace{1em}The advantage of boosting is that it can improve the model's stability through repeated training. In real life, many problems are highly complex. When making decisions with decision trees, processing a few features can lead to overfitting, reducing adaptability. Therefore, we need to use boosting methods to gradually build a Random Forest model, thereby improving its stability.

\begin{figure*}[t]
    \centering
    \includegraphics[width=0.6\linewidth]{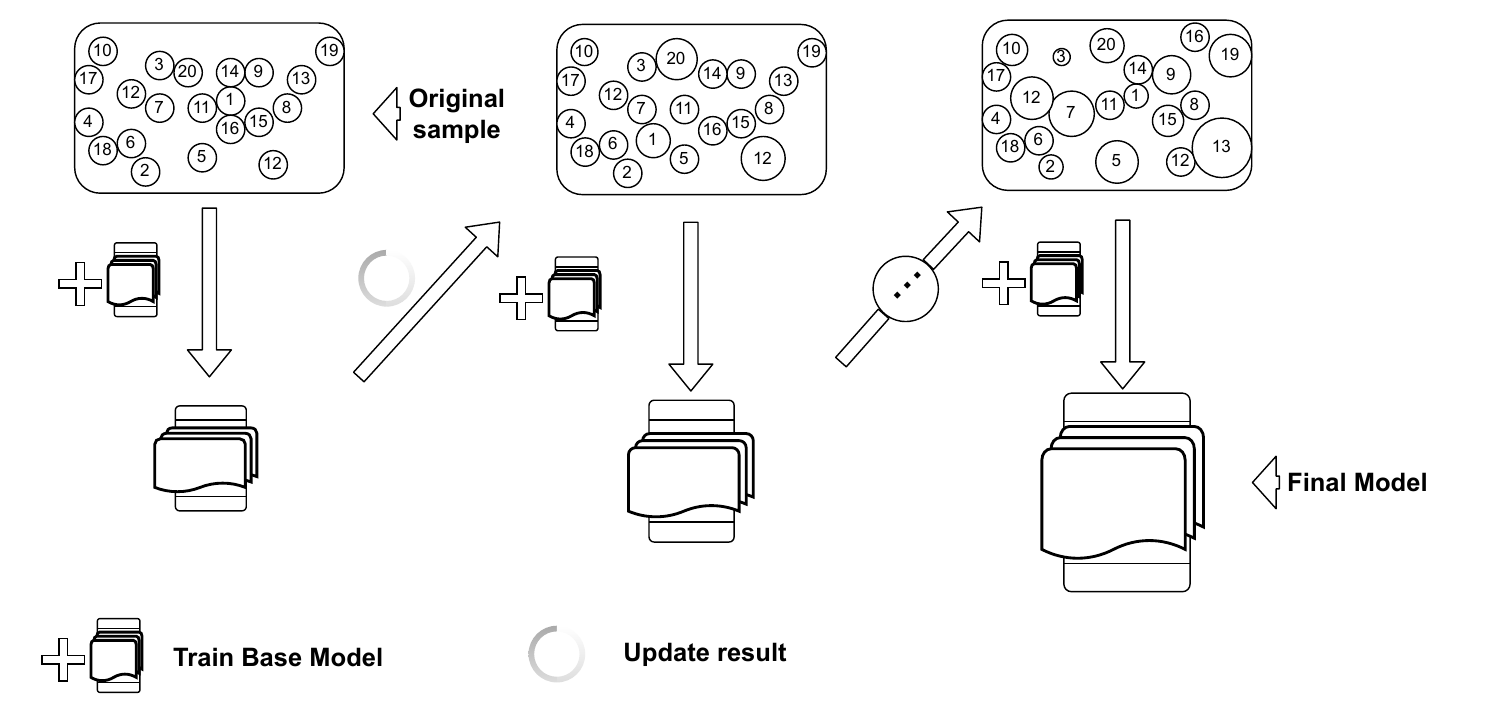} 
        \caption[Figure 17]{Boosting Detailed thought process diagram}
    \label{framework}
\end{figure*}

\subsubsection{Individual Learners}
An individual learner refers to a basic learner trained on a given dataset. It is the fundamental element of an ensemble model. It can be any type of learning algorithm, such as a support vector machine. Each individual learner is trained on a portion of the given data, generating corresponding classifiers or regressors, which are then combined to form a more powerful overall model.

In homogeneous ensembles, individual learners are called base learners, and all learners are of the same type.

In heterogeneous ensembles, individual learners are called component learners, and the learners include other types. Individual learning often refers to a single learner, while ensemble learning typically involves the integration of multiple learners in some way. Below is an example.

\begin{figure*}[t]
    \centering
    \includegraphics[width=0.6\linewidth]{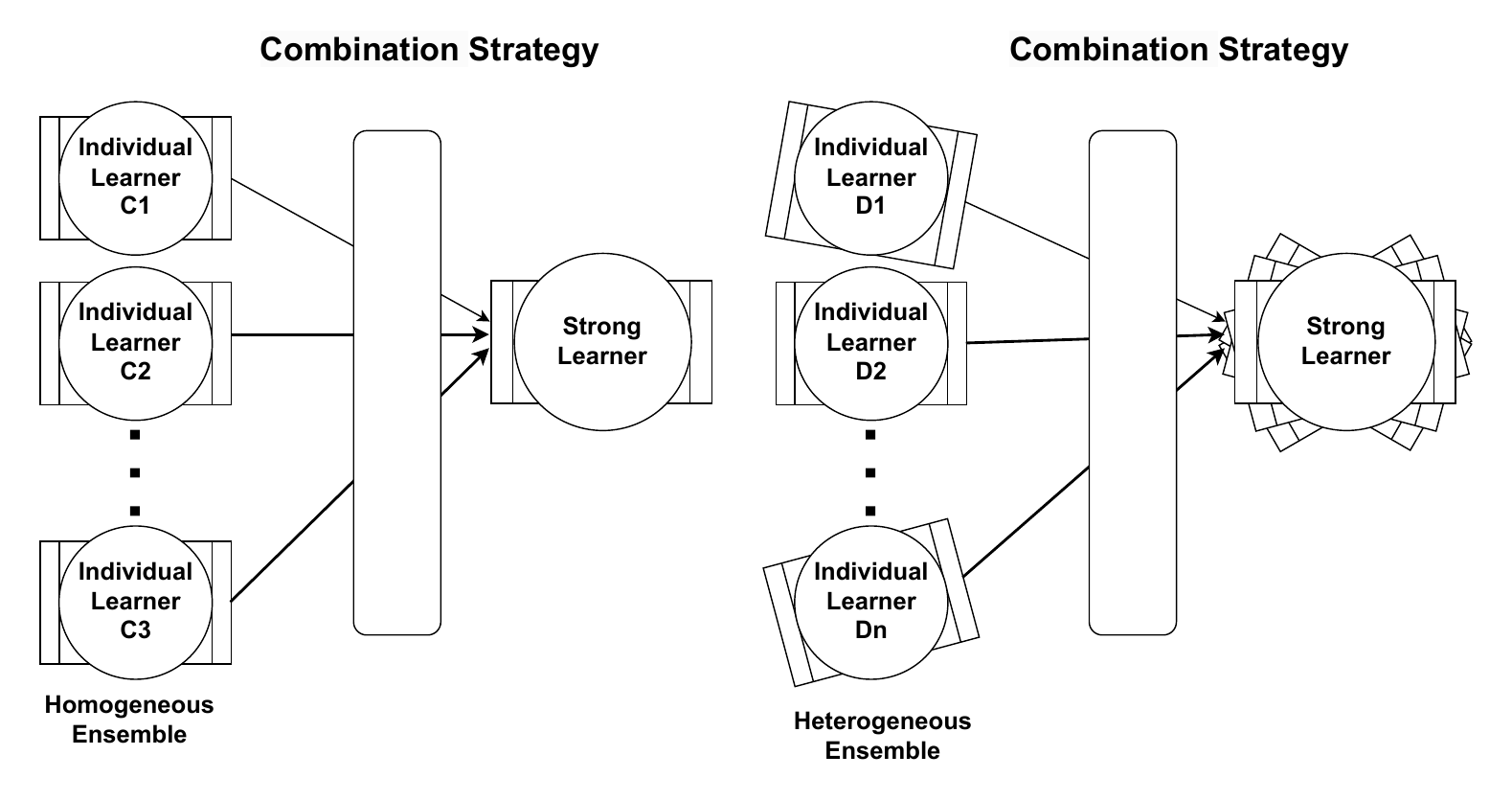} 
        \caption[Figure 18]{The two approaches to generating strong learners}
    \label{framework}
\end{figure*}

\subsubsection{Core Issues in Ensemble Learning}

The core issues in ensemble learning are divided into two aspects: the selection of learners and the construction of strategies.

\textbf{(1) Which Learners to Use?}

When selecting individual learners, we require:
\begin{itemize}

\item The performance of individual learners should not be too poor; they should not make significant errors.

\item There should be a certain level of difference between individual learners; they should not be almost identical.
\end{itemize}
When individual learners are too weak, the performance of the ensemble learning model will decline. The weaknesses of individual learners manifest as low accuracy, high bias, or insufficient learning ability for samples. Therefore, we need to avoid this situation and strive to increase the diversity of the model. Avoid similar learning effects that could affect the model's stability. 

Thus, selecting individual learners in ensemble learning becomes a significant challenge. We need to ensure both diversity and stable predictions. For problems not encountered by one learner, another learner should provide some complementary capabilities. By carefully selecting learners, we can ensure reasonable and effective learning outcomes, improving the ensemble learning model's ability to handle general problems and avoid overfitting.

\textbf{(2) Which Strategies to Use?}

To select appropriate combination strategies to build strong learners, there are two methods. Both methods are widely used in ensemble learning and have different characteristics and advantages. 

Below is an introduction to them.
\begin{itemize}

\item \textbf{Parallel Combination Methods:}Bagging: Constructs a large number of individual learners using the bootstrap method and averages or votes based on the results. It is suitable for large datasets with relatively simple individual learners that are not prone to overfitting.

Random Forest: Suitable for high-dimensional data or data with a large number of features.

\item \textbf{Traditional Combination Methods:}Boosting: Trains learners and continuously adjusts the process to account for previous errors. It is suitable for individual learners with small bias but high variance.

Stacking: First, multiple individual learners (of different types) are stacked together, then trained, and combined with a meta-learner to predict results. This method allows for better utilization of diverse learners. \citep{togatoropa2022optimizing}
\end{itemize}

\textbf{(3)When dealing with specific problems?}
\begin{itemize}

\item \textbf{ Data volume:} For example, whether the data volume is too large, whether it involves multiple directions, and whether it has specific impacts on other factors.

\item \textbf{ Individual learners:} Generally, the differences between learners should be considered, such as whether the learners are stable and have high accuracy.

\item \textbf{ Algorithm:} If the algorithm requires significant computational resources, we need to consider whether the chosen ensemble method can be better applied  and whether it meets the requirements of the combination strategy.\citep{li2021hybrid}
\end{itemize}
\hspace{1em}Finally, the model's response to abnormal problems should also be considered, such as how to better utilize noise in real-life scenarios. How to interpret high-dimensional data and use appropriate methods to ensure a clear and accurate understanding. After completion, how to evaluate the model, such as through comparative analysis and experimental verification.

\subsection{Algorithmic Approach}
After implementing decision trees, the Random Forest algorithm can be divided into three main aspects: drawing equal-sized samples, randomly selecting features, and building multiple trees.
The Random Forest model is based on the idea of bagging, using CART decision trees on learners to optimize the model. The approach is as follows:

In the preparation phase, we first draw part of the training data for training. Since the selection is random, it ensures that the decision tree samples have differences, providing the basic conditions for subsequent steps.

To ensure the stability of the decision tree quality, we perform sampling with replacement. During training, each tree may use part of the data from other trees, although some data may not be used.

For the possible scenario where a small number of samples are not used, we address this by implementing a large number of trees and using them as test models.

In summary, we first select data, draw N samples, and obtain N sample sets to train and produce initial results. This process is repeated in the second round.

\begin{figure*}[t]
    \centering
    \includegraphics[width=0.6\linewidth]{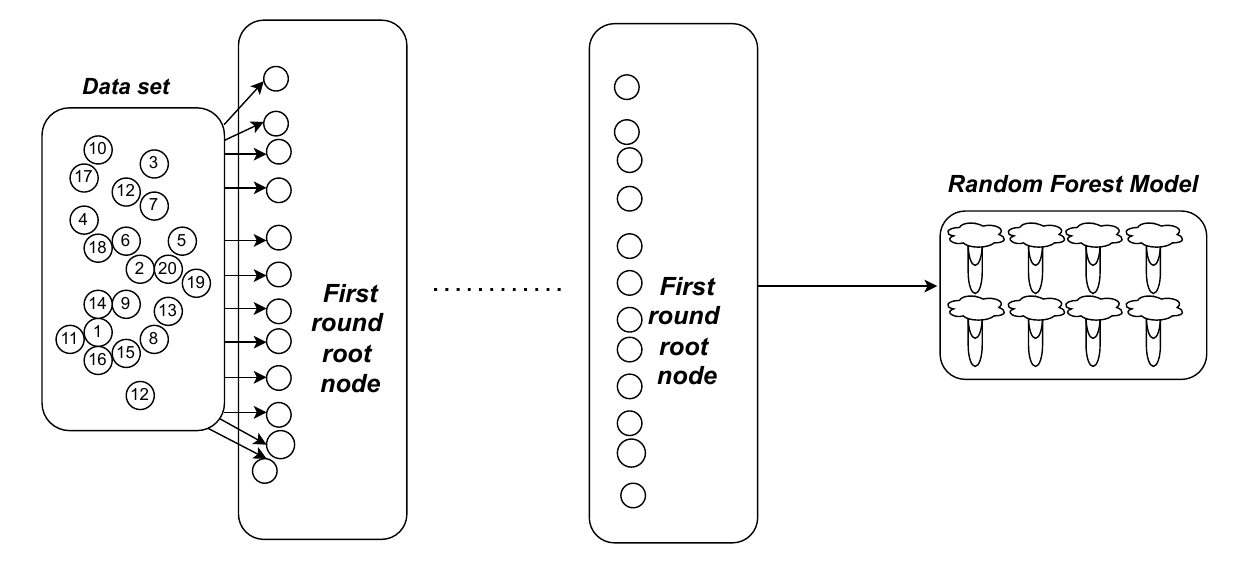} 
        \caption[Figure 19]{Drawing Equal-Sized Samples}
    \label{framework}
\end{figure*}

During model training, we randomly select features to ensure that different decision trees use as diverse features as possible.
Combined with the previous random sampling of data, these two aspects together improve the model's adaptability and accuracy.

Below are some advantages of randomly selecting features.
\begin{itemize}

\item \textbf{Low similarity:} By selecting different features, different decision trees have more options. Using all features for a single criterion can lead to redundancy in decision trees.

\item \textbf{Key features are prominent:} When data has high dimensionality, some features may not be very useful. Random selection helps exclude these features, enhancing the influence of key features. After training, key features become more prominent, making decision trees more effective.

\item \textbf{ Controllable complexity:} If we do not control the number of features used, many features may be calculated, often wasting computational resources. By controlling features, we can manage the model's complexity.
\end{itemize}
\hspace{1em}In summary, by randomly selecting features, we effectively solve the problem of decision tree homogeneity, enhance and strengthen key features, and control the model's computational resource consumption. This balances model computational complexity and applicability, ensuring its value and effectiveness in real-world problem-solving. Below is an example diagram of feature selection iteration.

\begin{figure*}[t]
    \centering
    \includegraphics[width=0.6\linewidth]{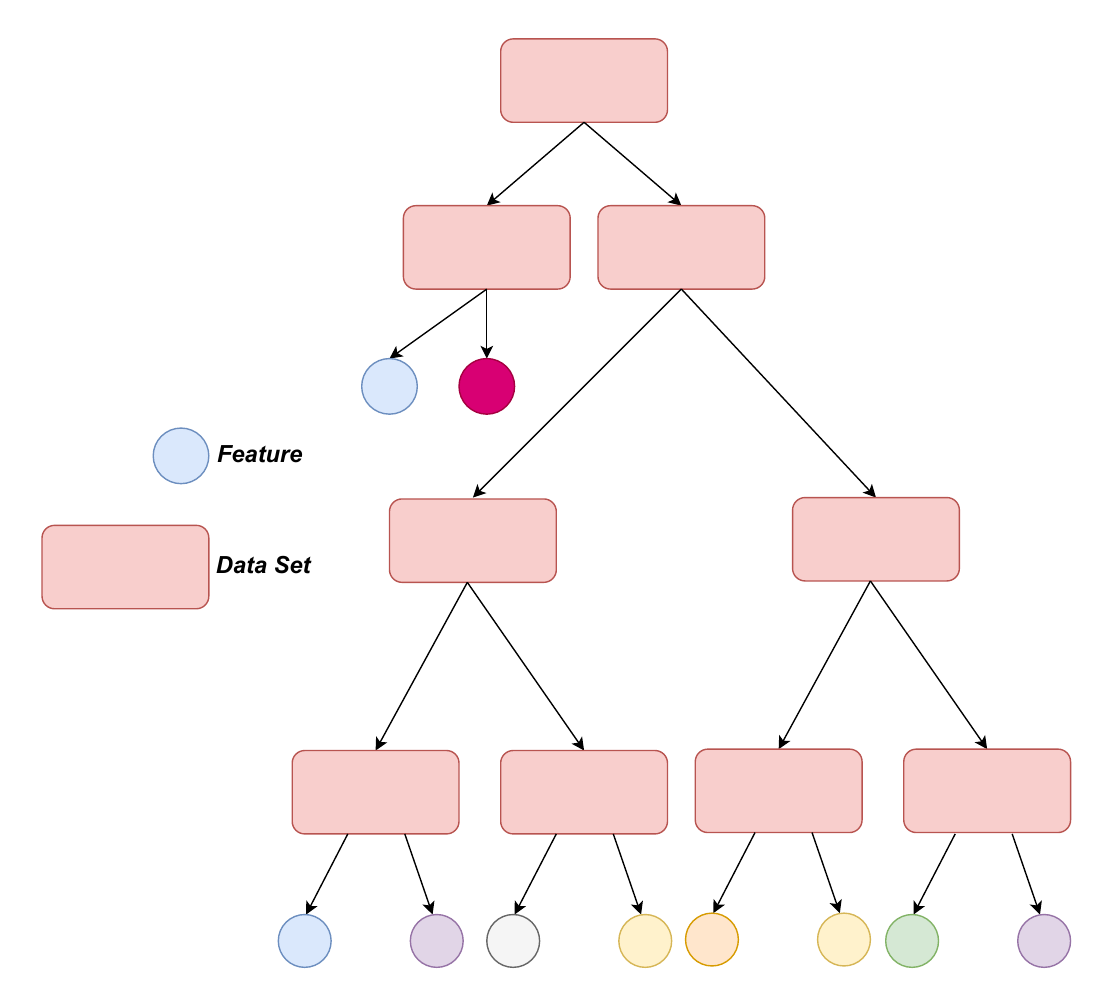} 
        \caption[Figure 20]{Feature Selection Iteration}
    \label{framework}
\end{figure*}

Through the previously mentioned steps of drawing equal-sized samples and randomly selecting features, we can perform a round of Random Forest training.

First, we draw samples. We randomly draw a certain number of samples from the original training dataset, keeping the total number of samples equal. This ensures the model's diversity and controls its generalization ability to some extent.

During training on each sample set, features are not deliberately selected. Finally, we build a large number of decision trees and combine multiple learners, making the model highly applicable to both classification and regression problems.
\begin{figure*}[t]
    \centering
    \includegraphics[width=0.6\linewidth]{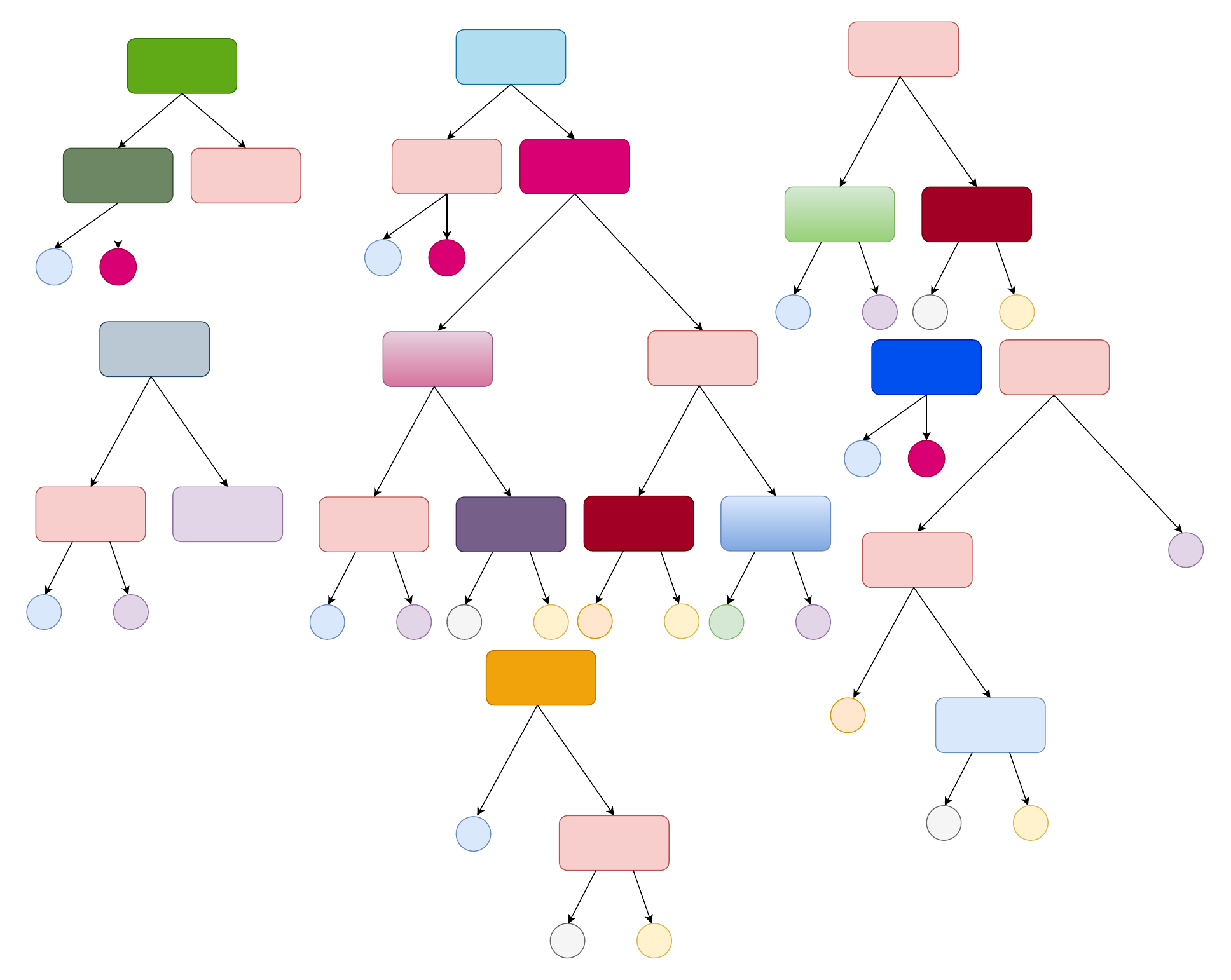} 
        \caption[Figure 20]{Feature Selection Iteration}
    \label{framework}
\end{figure*}

\subsection{Model Advantages and Disadvantages}
\subsubsection{Advantages}
The Random Forest model does not require complex feature dimensionality reduction when dealing with high-dimensional data; it can directly use datasets with many features. In feature selection, it can calculate the approximate weight of each feature, rank them, and control the overall structure. Additionally, it can create new features, further improving the model's generalization ability.

Random Forest uses parallel integration, effectively controlling overfitting during training. Its engineering implementation is simple, and training is fast, showcasing its advantages when dealing with large datasets. For the data imbalance issue in this paper, it also performs well.

Finally, Random Forest shows strong robustness to missing features, maintaining high prediction accuracy. In summary, the Random Forest model is a powerful and comprehensive machine learning algorithm suitable for handling high-dimensional, dense data. \citep{dong2020survey}

\subsubsection{Disadvantages}
When facing noisy data, the adaptability of Random Forest is still limited. Although it can overcome some issues by randomly selecting features and datasets, problematic datasets can still be collected into the corresponding learners during training, affecting overall decision-making and model stability.

Since Random Forest generates a large number of decision trees, compared to general decision tree problems, it requires more explanation. Decision trees vary in shape, and some may have obvious issues. How to provide reasonable and effective explanations, identify problematic points in decision trees with many issues, and offer solutions a challenges for model designers.

Therefore, better handling of abnormal problems and improving the model's interpretability require further consideration and processing. (Usually, careful discrimination is needed in data processing.) Additionally, for problematic decision trees, providing reasonable and effective explanations, analyzing the causes of problems, and verifying and correcting them are necessary to improve the model's interpretability. \citep{li2022benchmark}

\subsection{Parameter Tuning}
Our Random Forest model has five main parameters: maximum number of features, number of trees, maximum depth, minimum samples required to split an internal node, and minimum samples required at a leaf node. Initially, we do not know where to apply these parameters to achieve optimal results, which requires extensive processing and experimentation. This is where parameter tuning comes into play. Through parameter tuning, we can improve the model's robustness and accuracy in handling complex problems, ensuring its normal operation on general issues. \citep{wang2016novel}

The maximum number of features is a crucial parameter in the Random Forest model, corresponding to the step of randomly selecting features during model construction. Problems arise when this number is too large or too small.

When the maximum number of features is too small, the number of features available for a decision tree decreases. In an extreme case, if a decision tree has only one feature, its applicability will significantly decrease, clearly not meeting the requirements. The diversity of decision trees will also decrease, inevitably reducing the model's generalization ability.

However, when the maximum number of features is too large, problems also arise. If it is too large, decision trees will use the same features, weakening their ability to handle noisy data and leading to overfitting.

In general, the maximum number of features should account for 50\%-75\% of the total features. However, this is not absolute, and testing for both larger and smaller cases is necessary to ensure the model's generalization ability.

The number of trees ensures that after randomly selecting data and features, all data and features are better covered. A small number of trees can lead to the following problems:
\begin{itemize}

\item \textbf{Underfitting:} If there are too few trees, such as 10, many scenarios cannot be covered. When the data volume is large, it becomes impossible to handle complex logical relationships in the data, leading to underfitting.

\item \textbf{Poor adaptability:} In real-life problems, many scenarios require extensive experimentation to simulate. With fewer trees, the predictive ability weakens, inevitably reducing adaptability in real-life problem-solving.

\item \textbf{High variance:} Since the Random Forest model is based on two types of randomness, the disadvantage of randomness is the incomplete consideration of problems. If the number of trees is not increased, the disadvantage of randomness will be exposed, leading to unstable performance when the model is applied to different datasets.
\end{itemize}
\hspace{1em}A large number of trees can also cause the following problems:
\begin{itemize}

\item \textbf{Overfitting:} When there are too many trees, similarity issues become prominent. Over-consideration may capture subtle features in the dataset that are not useful in practical applications, leading to overfitting.

\item \textbf{Imbalanced benefits:} Using too many trees means considering all possible scenarios. When the data volume is large, significant resources are consumed without a corresponding improvement in results, making it inefficient.
\end{itemize}
\hspace{1em}In general, the number of trees should be controlled between 50 and 100. Too many or too few trees can cause problems.

The maximum depth of decision trees has a significant impact on the complexity of the Random Forest model. Both too large and too little depths can cause a series of problems.
\begin{itemize}

\item When the maximum depth is too large, the model may undergo repeated training, considering too many scenarios. Subtle features specific to the dataset may be included, causing unnecessary issues, such as overfitting.

\item When the maximum depth is too small, decision trees become too simple, making it difficult to handle complex problems. The model's generalization ability and accuracy cannot be guaranteed.
\end{itemize}
\hspace{1em}Therefore, we often carefully select the maximum depth based on actual situations, generally controlling it between 4 and 10.

In the Random Forest model, whether to split an internal node has a significant impact on the generation of decision trees, representing the minimum number of samples required to split a node.

First, setting an appropriate minimum number of samples is crucial for the performance and generalization ability of the Random Forest model. When it is too small, the threshold for secondary or multiple splits is lower. This means decision trees can more easily split nodes, but it may also add unnecessary processing for irrelevant parts. These differences may not be significant on the training set, but when the scenario changes, the lack of generalization ability becomes apparent.

Conversely, when it is too large, we cannot process data that needs further splitting, limiting the growth depth of decision trees, making them simpler, and reducing the risk of overfitting.

Generally, setting an appropriate minimum number of samples benefits the diversity of decision trees, ensuring their differences and improving the model's stability.

Typically, the minimum number of samples required to split an internal node is determined based on the data volume, with no fixed range.

\subsubsection{Minimum Samples Required at a Leaf Node}

The minimum number of samples required at a leaf node refers to the minimum number of samples needed to split a node again. Both too-large and too-small values significantly impact the model's generalization ability.

When the minimum number of samples at a leaf node is too small, such as 1, each leaf node may split further. If there are outliers in the model, they will be included in the decision tree, harming the model's generalization ability and increasing the risk of overfitting.

When the minimum number of samples at a leaf node is too large, the model may not consider many scenarios, making it unable to capture complex relationships.

Meanwhile, by setting the minimum number of samples for leaf nodes, we can regulate the model's complexity, which helps conserve computational resources. When appropriate leaf nodes are selected, the resulting decision tree is usually well-balanced—neither overly complex nor too simple. This also reduces obstacles encountered when interpreting the model.

\subsubsection{GridSearch Hyperparameter Tuning Demonstration}
GridSearch is a hyperparameter tuning algorithm we learned in our artificial intelligence and big data course. It allows us to control a model's complexity through hyperparameters, thereby influencing aspects such as the model's accuracy and training time. \citep{yates2023cross} 

To provide an intuitive understanding and analysis, we allocated five parameters. Through experimentation, we found that the minimum number of samples required to split an internal node and the minimum number of samples for a leaf node had little impact on the results after parameter settings (this is primarily related to the data volume). To improve generalization, we set these values to >1. Therefore, before tuning, we set the minimum number of samples to split an internal node to 2 and the minimum number of samples for a leaf node to 2. We ranked them by the size of cross-validation.  

Below are 3D visualizations of GridSearch hyperparameter settings for five groups, including risk-variable proxies. The color gradient ranges from light blue to blue, then to light red, and finally to dark red, with darker colors indicating better performance.  

\begin{figure*}[t]
    \centering
    \includegraphics[width=0.6\linewidth]{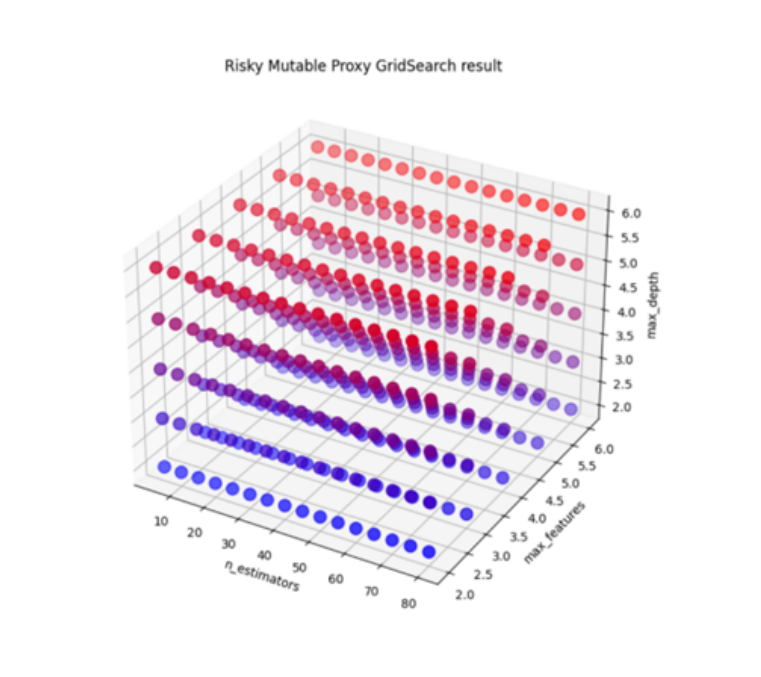} 
        \caption[Figure 20]{Risky Mutable Proxy}
    \label{framework}
\end{figure*}

The above diagram detail the processing effects of different parameter selections for various vulnerabilities. Taking the last example of common loss, the poorer-performing points are mainly concentrated around a very small maximum depth (2) (underfitting issue). As the maximum depth increases, the processing effect improves. The best performance is observed in the range of maximum depths 3–4, with performance declining as depth increases further (corresponding to overfitting issues).  

Through the visual analysis above, we adjusted the parameter settings for different vulnerabilities, with the results shown in the following table:  

\begin{table}[h]
    \centering
    \setlength{\abovecaptionskip}{0.cm}
    
    \resizebox{0.8\linewidth}{!}{
    \begin{tabular}{cccccccc}
    \toprule
        \textbf{Random Forest Name} &\textbf{Maximum Features} & \textbf{Number of Trees}  & \textbf{Tree Depth}  \\  \hline
        Risk Mutable Proxy &3  &50  & 4  \\
        ERC-721 Reentrancy &4  &50  &5 \\
        Unlimited Mining &4  &75  & 4 \\
        Missing Requirements &4  &50  &4 \\
        Public Burn &5  &55  & 3\\
        \hline
    \end{tabular}}
    \caption{GridSearch setting}
    \label{performance_measures}
\end{table}

\subsection{Random Forest Results for Each Vulnerability  }

Using GridSearch hyperparameter tuning, we selected the best-performing decision tree from the generated options. Below is a detailed explanation of the parameters.  

Taking the optimal solution for the risk-variable proxy as an example:  
A3 represents the feature value (A3 <= 0.5 means A3 = 0, indicating the feature is not satisfied).  Gini is the Gini coefficient (smaller values indicate higher purity).  Samples refers to the number of samples corresponding to the feature. Value divides the data into two categories: the first class is 0, and the second is 1, representing whether the feature is unsatisfied or satisfied, respectively.  

\subsection{Conclusion}  
Through the transition from decision trees to random forests, the model has undergone a comprehensive leap. On one hand, I have a complete process from raw data to the final decision tree generation, along with data from each stage, resulting in strong interpretability.  

Additionally, during the construction of the random forest, I conducted extensive visualization work for the selection of five key parameters. By utilizing the GridSearch hyperparameter tuning method, significant efforts were made to provide a more intuitive understanding and insight into the impact of parameters across various vulnerabilities.

\appendix
\bibliographystyle{ACM-Reference-Format}
\bibliography{sample-base}

\end{document}